\documentclass{aa}

\usepackage[dvips]{graphicx}
\usepackage{amsmath}
\usepackage{amssymb}

\newcommand{\Teff}{\mbox{$T_{\mbox{\rm \tiny eff}}$}}

\newcommand{\Phoenix}{\mbox{\tt PHOENIX}}
\newcommand{\Marcs}{\mbox{\tt MARCS}}

\begin{document}

\title{Near-IR Spectra of Red Supergiants and Giants}
\subtitle{I. Models with Solar and with Mixing-Induced Surface Abundance Ratios
\thanks{Selected theoretical spectra (see text) can be retrieved in FITS format 
at CDS via anonymous ftp to cdsarc.u-strasbg.fr (130.79.128.5), or via
http://cdsweb.u-strasbg.fr/cgi-bin/qcat?J/A+A/}}

\author{A. Lan\c{c}on\inst{1}
     \and
        P.H. Hauschildt\inst{2}
     \and 
	D. Ladjal\inst{1,3}
     \and
        M. Mouhcine\inst{4} }

\offprints{A. Lan\c{c}on \email{lancon@astro.u-strasbg.fr}  }

\institute{
 Observatoire Astronomique de Strasbourg,
 Universit\'e L.\,Pasteur \& CNRS (UMR 7550), Strasbourg, France
\and
 Hamburger Sternwarte, Gojenbergsweg 112, 21029 Hamburg, Germany
\and
 Institute of Astronomy, Katholieke Universiteit, Celestijnenlaan 200\,B, 
 3001 Leuven, Belgium
\and
 Astrophysics Research Institute, Liverpool John Moores University, 
 Twelve Quays House, Egerton Wharf, Birkenhead, CH41~1LD, UK 
}

\date{Received July 2006 / Accepted 2 January 2007}

\authorrunning{A. Lan\c{c}on et al.}

\abstract
{
It remains difficult to interpret the near-IR emission of young
stellar populations. One main reason is our incomplete understanding
of the spectra of luminous red stars.
}
{
This work provides a grid of theoretical spectra of red giant and 
supergiant stars, that extends through optical and near-IR 
wavelengths.  For the first time, models are also provided with modified
surface abundances of C, N and O, as a step towards accounting
for the changes that occur due to convective dredge-up in red
supergiants or may occur at earlier evolutionary stages in the
case of rotation. The aims are (i) to assess how well current models
reproduce observed spectra, in particular in the near-IR, (ii) to
quantify the effects of the abundance changes on the spectra, and
(iii) to determine how these changes affect estimates of fundamental
stellar parameters.
}
{
Spectra are computed with the model atmosphere code \mbox{\tt PHOENIX} 
and compared with a homogeneous set of observations.
Although the empirical spectra have a resolution of only
$\lambda/\Delta\lambda \sim 1000$, we emphasize that models must
be calculated at high spectral resolution in order to
reproduce the shapes of line blends and molecular bands.
}
{
Giant star spectra of class III can be fitted extremely well
at solar metallicity down to $\sim$3400\,K, where difficulties appear
in the modelling of near-IR H$_2$O and TiO absorption bands.
Luminous giants of class II can be fitted well too,
with modified surface abundances preferred in a minority of cases,
possibly indicating mixing in excess of standard first dredge-up.
Supergiant stars show a larger variety of near-IR spectra, and
good fits are currently obtained for about one third of the observations only.
Modified surface abundances help reproducing strong CN bands, but 
do not suffice to resolve the difficulties. 
The effect of the abundance changes on the estimated \Teff\ depends on the
wavelength range of observation and can amount several 100\,K.
}
{
While theoretical spectra for giant stars are becoming very
satisfactory, red supergiants require further work. The model grid
must be extended, in particular to larger micro-turbulent velocities.
Some observed spectra may call for
models with even lower gravities than explored here
(and therefore probably stellar winds),
and/or with more extreme abundances than predicted by
standard non-rotating evolution models. Non-static atmospheres models
should also be envisaged.
}
 

\maketitle

\section{Introduction}

Red supergiants and red giants are the most luminous stars in, respectively,
star forming or old passive galaxies. Being cool, they are the dominant
sources of near-IR light. In highly
reddened starburst galaxies, the near-IR light from red supergiants
is sometimes the only direct information available on the stellar populations.
Models for their spectra are thus important, even though they
are also particularly difficult to construct (rich molecular line spectrum,
extended atmospheres). If they are 
successful in reproducing empirical spectra, it will be legitimate
to use them instead of observed spectral libraries in future analyses
of galaxies. 

Recently, Levesque et al. (2005) have shown that up-to-date models
compare well with optical observations of red supergiants, 
and shown that this success helps in explaining the
location of observed red supergiants in the HR-diagram.
At near-IR wavelengths (1--2.5\,$\mu$m), the most prominent molecular
features are those of CO and CN. Their sensitivity to surface gravity 
and effective temperature (\Teff ) has been the basis of the 8-colour
classification system of White \& Wing (1978), although they are 
also sensitive to other parameters (Tsuji 1976, McWilliam \& Lambert 1984,
McGregor 1987,
Bessell et al. 1991, Origlia et al. 1997). Strong CN bands are 
predicted for low gravity stars with temperatures around 4500\,K, and
are indeed observed in local red supergiants (Lan\c{c}on \& Wood 2000) and 
in extragalactic objects such as the bright star clusters of M\,82
(Lan\c{c}on et al. 2003). However, it had not been verified until now whether
models are capable of reproducing the strengths of various CN and CO bands
throughout the near-IR range simultaneously. Nor whether they can match
optical and near-IR properties together.

An important aspect not accounted for in recent collections of model spectra
for red supergiants is internal mixing. 
Standard stellar evolution predicts non-solar
surface abundance ratios due to convective dredge-up in the red supergiant 
phase (Iben 1966, Maeder 1981). 
Early observations had pointed out the inadequacy of
solar abundance ratios in individual cases (e.g. $\alpha$ Ori, Beer et 
al. 1972). More recently, both theory and observations showed
that main sequence rotation or other processes are capable of mixing 
CNO-cycle products into the atmosphere even before the red supergiant phase 
is reached (Maeder \& Meynet 2001, Trundle \& Lennon 2005). 
In red supergiants, He and $^{14}$N surface abundances
are typically enhanced while $^{16}$O and $^{12}$C abundances are reduced. 
Modified abundances of C,N and O alter the relative strengths of the 
predominant molecules. 

In this paper, we present recent \Phoenix\ models specifically computed
to address some of the above points. 
The emphasis is on the effects of
non-solar abundance ratios of C,N and O; a more complete study of other
parameters (in particular micro-turbulent velocities) has been started and
will be reported in a forthcoming paper. 
The model assumptions are described in
Sect.\,\ref{models.sec} and the predicted colours and molecular features 
in Sect.\,\ref{trends.sec}. In Sect.\,\ref{twocolour.sec} and 
\ref{spectra.sec}, the models are compared with spectroscopic 
data covering wavelengths from 1 to 2.4\,$\mu$m or, 
for a subsample, from 0.51 to 2.4\,$\mu$m. Giants of class III, 
luminous giants of class II and supergiants of class I are discussed
successively. The discussion in Sect.\,\ref{discussion.sec} focuses on 
fundamental parameter determinations from spectra, including the effects
of mass and surface abundances on \Teff . A brief conclusion summarizes
the results.

\section{\Phoenix\ models with solar and modified abundances}
\label{models.sec}

\subsection{Summary of model ingredients}

\begin{figure}
\includegraphics[clip=,width=0.49\textwidth]{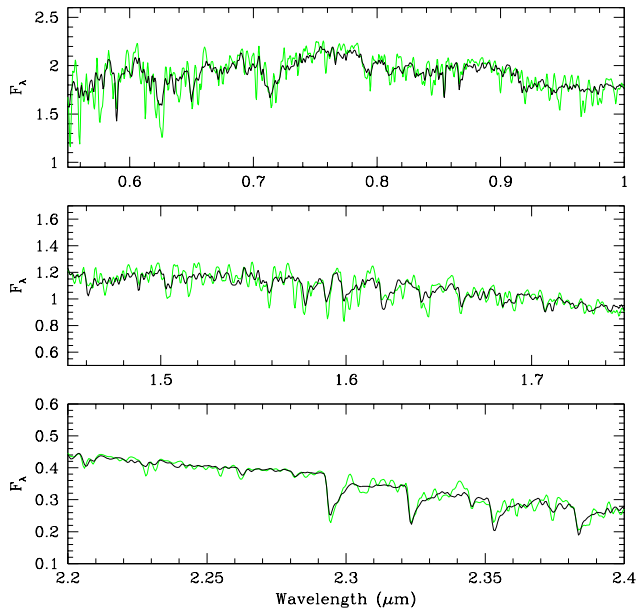}
\caption[]{Typical differences between \Phoenix\ 
spectra obtained with an initial 
wavelength sampling step of 2\,\AA\ (grey) and 0.1\,\AA\ (black). Both spectra
have been smoothed by convolution with a gaussian with FWHM=15\,\AA. The
models shown have T$_{\rm eff}$=4000\,K, log($g$)=1, M=1\,M$_{\odot}$, but
differences are important for any of the calculated models. Only the
high resolution calculations match the data.}
\label{effetRes.fig}
\end{figure}

The model atmospheres and synthetic spectra were computed with \Phoenix\ 
version 13.11.00B.
The model setup is identical
to that of Ku\v{c}inskas et al. (2005). We recall only the most relevant
details here. The models are computed in spherical symmetry.
A mixing length to pressure scale height ratio of 2.0 is used for
all models.
Dust is allowed to
form in the atmospheres but is assumed to immediately rain out of the 
photospheric layers; therefore, no dust opacities are used in the models
shown here. This is an important assumption for cool models with large
extensions. In addition, all models presented here have large
enough gravities to not produce a radiatively driven wind and,
therefore, winds are not included.

The model spectra were computed specifically for comparison with data
that has a spectral resolving power of order 1000, i.e. $\Delta \lambda \simeq 
10$\,\AA\ at 1\,$\mu$m. {\em Nevertheless, we emphasize
that the model spectra must be computed at high spectral resolution
before smoothing, in order to sample individual absorption line 
profiles properly and to obtain low resolution spectra that 
resemble observations}\ (Fig.\,\ref{effetRes.fig}). 
We used a wavelength sampling step of 0.1\,\AA\ throughout. 
Using only half these points produces negligible changes 
at $\lambda>8000\,\AA$ (0.1\,\% rms), and small variations in the shapes 
of the strongest optical bands at $\lambda<8000\,\AA$ (2\,\% rms).
The small sampling step used here is an important change with respect
to previous collections of \Phoenix\ spectra, which were computed with 
an initial wavelength sampling step of 2\,\AA\ (e.g. Ku\v{c}inskas et al., 2005,
and models included in the library of Martins et al., 2005).

The models discussed cover effective temperatures ranging from 2900
to 5900\,K and gravities in the set log($g$)=\{$-1$,$-0.5$,0,1,2\}
(cm.s$^{-2}$). 
The micro-turbulent velocity is set to
a constant value of $v_{\rm mic}$\,=\,2\,km\,s$^{-1}$, 
except in a few exploratory models. 
Values of 2 to 3\,km\,s$^{-1}$
are typical for red giant stars (Smith \& Lambert 1985). 
A more extensive grid of models covering higher values of this
parameter is in the process of being calculated.
Two stellar masses are considered\,: 1\,M$_{\odot}$ and 15\,M$_{\odot}$.
Models at 9\,M$_{\odot}$ were also computed, but the differences with 
the 15\,M$_{\odot}$ ones are negligible. For M=1\,M$_{\odot}$, many of the
calculations at log($g$)=$-1$ did not converge (radiation pressure incompatible
with the assumption of no winds),
and this mass-gravity combination is therefore excluded from 
the discussion. 
We also restrict the discussion to optical and near-IR wavelengths, with
a focus on wavelengths between 0.81 and 2.4\,$\mu$m.

\subsection{Abundances}

The reference set of models assumes solar abundances,  based on the review
of Grevesse \& Noels (1993). The values most relevant to our study are
summarized in Col.\,2 of Tab.\,\ref{abundances.tab}. A subset of 
models with solar-scaled abundances but $\log(Z/Z_{\odot})=-0.3$ was also
computed but will only be discussed briefly in Sect.\,\ref{Teffscales_models.sec}.

The second set of models has the same metallicity Z=0.02 as the reference 
set, but modified abundances of $^4$He, $^{12}$C, $^{14}$N, $^{16}$O
(Col.\,5 of Tab.\,\ref{abundances.tab}). 
In the following, the adopted modified 
abundances will be refered to as ``RSG-specific abundances".

The RSG-specific abundances were selected by 
inspection of the evolutionary tracks
of Schaller et al. (1992; their case of standard mass loss) 
for stars with initial masses above 7\,M$_{\odot}$, at evolutionary
timesteps with effective temperatures below 4500\,K. The values selected are 
representative of the final red supergiant stages of a star of
initial mass 20\,M$_{\odot}$ (\Teff$\simeq$3550\,K). 
Stars of lower initial masses would 
have RSG abundances closer to the main sequence values, while 
the tracks at 25\,M$_{\odot}$ reach significantly larger modifications
(tracks above 25\,M$_{\odot}$ don't extend to the low effective temperatures
of red supergiants).
Note that the initial mass fractions of Schaller et al. (1992) are 
not exactly the same as assumed in our reference set (mainly because
of their larger He abundance), but that these differences are small
compared to those that distinguish red supergiants from zero age main sequence
stars.

\begin{table*}
\begin{center}
\caption[]{Surface abundances (by mass)}
\label{abundances.tab}
\begin{tabular}{c|lll|lll}
        & Adopted   & Geneva   & Padova & Adopted  & Geneva & Padova \\
        & reference & 1992-1994 & 1993 & RSG-specific & 1994 & 1993  \\
Element &   set     &  ZAMS    &   ZAMS & set    & RSG   & RSG    \\ 
(1)  &  (2)  & (3)  & (4)  & (5)  & (6)  \\ \hline
$^1$H    & {\bf 0.703} & 0.680 & 0.700 & {\bf 0.580}   & 0.55  & 0.63 \\
$^4$He   & {\bf 0.280}  & 0.300 & 0.280 & {\bf 0.400}   & 0.43  & 0.35 \\
$^{12}$C & {\bf 0.0030}  & 0.0044 & 0.0049 & {\bf 0.0022 } & 0.0020 & 0.0032 \\
$^{14}$N & {\bf 0.00092}  & 0.0014 & 0.0012 & {\bf 0.0065} & 0.0080 & 0.0051 \\
$^{16}$O & {\bf 0.0083}  & 0.0106 & 0.0106 & {\bf 0.0076} & 0.0068 & 0.0084 \\
\hline
\end{tabular} 
\end{center}
{\em Notes:}\ 
{\bf Col. 2:} Abundances adopted in our reference set of solar metallicity 
models.
{\bf Col. 3:} For comparison, main sequence abundances of Schaller et al. 
(1992), also used by Meynet et al. (1994). Metal abundance ratios based on 
Anders \& Grevesse (1989).
{\bf Col. 4:} Main sequence abundances of Bressan et al. (1993). Metal 
abundance ratios based on Grevesse (1991).
{\bf Col. 5:} Abundances adopted in our RSG-specific set of models (see text).
{\bf Col. 6:} Final RSG abundances of Meynet et al. (1994) 
for 20\,M$_{\odot}$ stars.
{\bf Col. 7:} Final RSG abundances of Bressan et al. (1993) for 
20\,M$_{\odot}$ stars.  
\end{table*}

In Tab.\,\ref{abundances.tab}, the adopted abundances are compared 
to other values in the literature. The tracks of Meynet et al. (1994)
assume larger mass loss rates than those of Schaller et al. (1992).
More ingredients distinguish the models of Bressan et al. (1993) from those
of the Geneva group. Nevertheless, predicted surface abundance alterations 
are of a comparable amplitude. Evolutionary tracks for stars with initial
rotation predict that comparable $^{14}$N enhancements are reached 
already by the end of the main sequence (Meynet \& Maeder 2000, 2003), and
are further increased during the RSG phase. The achieved abundance ratios
depend strongly on initial rotation velocity, on initial mass and 
on mass loss prescriptions. 
One major difference between rotating and non-rotating models is 
the length of time a star spends as a red supergiant
with modified abundances in one and the other case
(see also Maeder \& Meynet 2001). 

Mixing also occurs along the red
giant branch and asymptotic giant branch 
for low and intermediate mass stars.
The RSG-specific abundances adopted here are more extreme than those 
obtained through 1st dredge-up on the RGB (e.g. Iben 1964,
Charbonnel et al. 1996, Girardi et al. 2000). 
In particular, the RSG-specific enrichment in He and $^{14}$N 
and the drop in H and $^{12}$C are larger than expected from 1st dredge-up. 
More mixing may however occur through
second dredge-up on the early asymptotic giant branch (for stars with 
initial masses above about 4\,M$_{\odot}$, e.g. Becker \& Iben 1979,
Boothroyd \& Sackmann 1999), and though ``non-standard"
extra mixing for low mass stars that have evolved on the red giant branch past 
the RGB-bump (Charbonnel 1994, Charbonnel \& do Nascimento 1998).
Both these processes affect relatively luminous giant stars. The second
one seems to be less efficient at the quasi-solar metallicities we consider
than for population II stars. Therefore, we expect our RSG-specific 
abundances {\em not} to be appropriate for most solar neighbourhood giants of 
class III, while they might be relevant to some giants of class II.

We note that future calculations with modified surface abundances will
include mixing-induced enhancements in the $^{13}$C abundance, since 
$^{13}$CO is a clear feature in near-IR spectra of cool stars.
The effects of recent changes in measurements of the solar abundance 
ratios (Asplund et al. 2005) will also be investigated.

\subsection{Spectra in numerical form}
The model spectra for M=1\,M$_{\odot}$ with solar abundances,
for M=15\,M$_{\odot}$ with solar abundances, and 
for M=15\,M$_{\odot}$ with RSG-specific abundances,
are made available in FITS format through 
CDS.
Because the quality assessments made in this paper are restricted
to resolutions of order $10^3$ in the near-IR (and a few hundred 
at wavelengths below 0.81\,$\mu$m), the spectra made available 
are smoothed with a Gaussian to a full width at half
maximum of 2\,\AA. The initial models, calculated with a 
wavelength step of 0.1\,\AA, can be requested from A.L. or P.H.H. 

\section{Trends in the models} 
\label{trends.sec}

\begin{figure*}
\includegraphics[clip=,width=\textwidth]{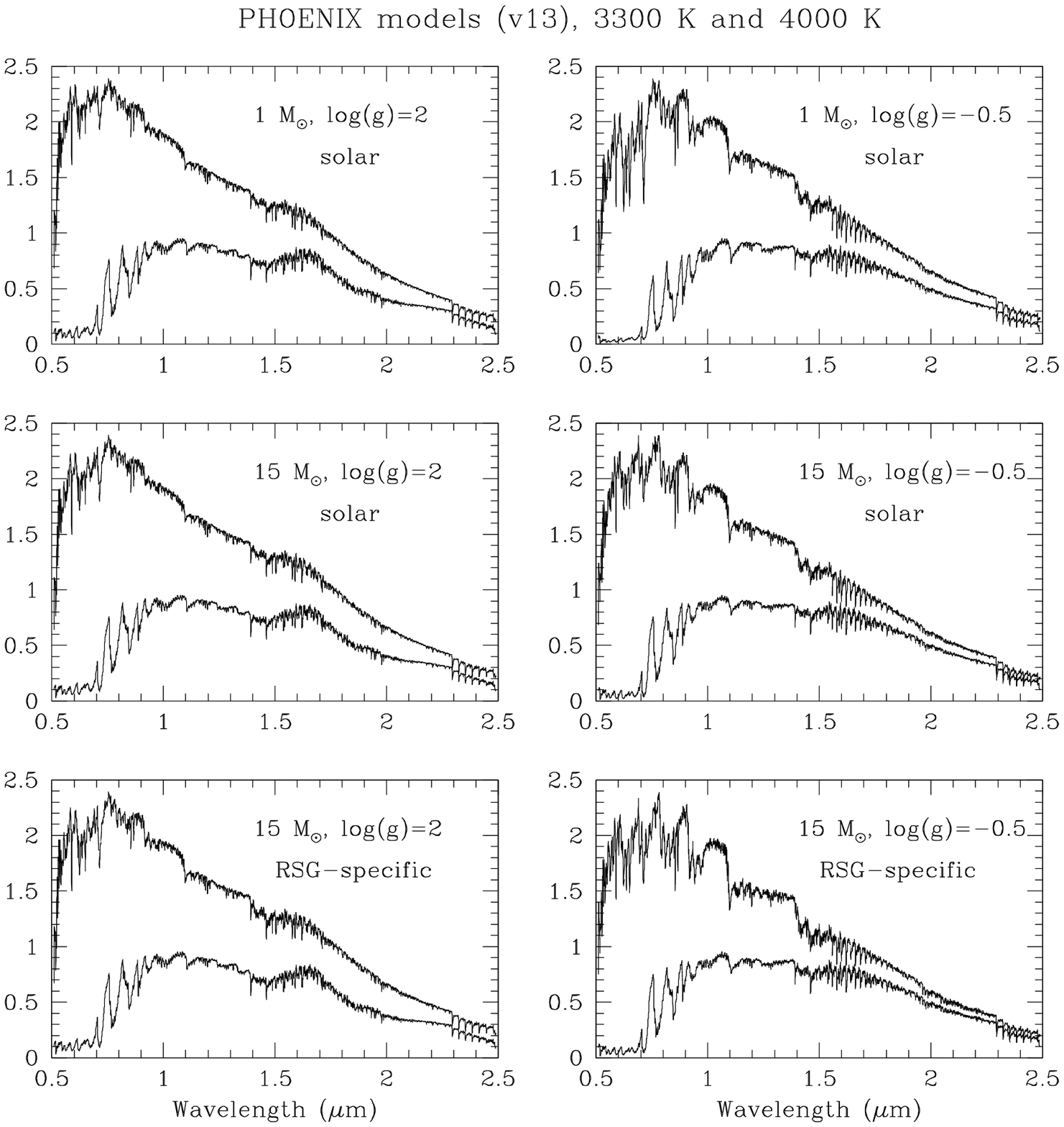}
\caption[]{Effects of gravity, temperature and surface abundances on
\Phoenix\ model spectra. In each figure, the upper spectrum has
$\Teff = 4000$\,K and the lower one $\Teff = 3300$\,K. For easier
comparison, fluxes have been normalized to values comparable to
those in the upper left diagram. The figures on the left are at log($g$)=2,
those on the right at log($g$)=-0.5. The upper figures
are for solar abundances, the lower ones for RSG-specific abundances.
The effect of mass (1\,M$_{\odot}$ vs. 15\,M$_{\odot}$) is too small
to be easily identified on this type of figure.}
\label{modeltrends.fig}
\end{figure*}

Spectra illustrating the effects of mass, gravity and surface abundances
are provided in Fig.\,\ref{modeltrends.fig}. In this 
section, we will discuss quantitative trends using selected
colours and molecular band indices. The indices
are measured for each spectrum using:
(i) the standard J, H, K filter passbands of Bessell \& Brett (1988);
(ii) narrow and intermediate-band filters as described in
Tab.\,\ref{indexdef.tab}. 
All narrow and intermediate filter passbands are approximated with rectangles
of identical central wavelength and width as the filters in the original
references (as already done by Bessell et al. 1989). 
A model Vega spectrum provides
the zero points in all passbands. 

\begin{table}
\caption[]{Filter and index definitions}
\label{indexdef.tab}
\begin{tabular}{llll}
Filter & Center  & Width & Notes \\
       & ($\mu$m)& (\AA) & \\ \hline
104 & 1.0395 & 50 & quasi-continuum (1) \\
108 & 1.0800 & 60 & CN (1) \\
110 & 1.1000 & 50 & quasi-continuum near CN (1) \\
220 & 2.2000 & 1100 & quasi-continuum (2) \\
236 & 2.3600 & 800  & 1st overtone CO (2) \\
COH & 1.6222 & 80  & 2nd overtone CO \\
COHc1 & 1.6160 & 30 & absorption-minimum near CO \\
COHc2 & 1.6285 & 30 & absorption-minimum near CO \\
\hline
Index & \multicolumn{3}{l}{Definition (3)} \\ \hline
104-220 & \multicolumn{3}{l}{$-2.5\log$(104/220)$+$cst}  \\
CO(2.3) & \multicolumn{3}{l}{$-2.5\log$(236/220)$+$cst}  \\
CO(1.6) & \multicolumn{3}{l}{$-2.5\log$[2\,COH/(COHc1+COHc2)]$+$cst}  \\
CN(1.1) & \multicolumn{3}{l}{$-2.5\log$(110/108)$+$cst}  \\
\hline
\end{tabular}
\\
{\em Notes :}\ (1) Adapted from the 8-colour system of R.F.\,Wing
(White \& Wing 1978). (2) Adapted from Frogel et al. (1978). 
(3) cst stands for a constant that gives the index the value 0 for 
a model spectrum of Vega.
\end{table}

\subsection{Colours}

\begin{figure*}
\centerline{
\includegraphics[clip=,angle=270,width=0.48\textwidth]{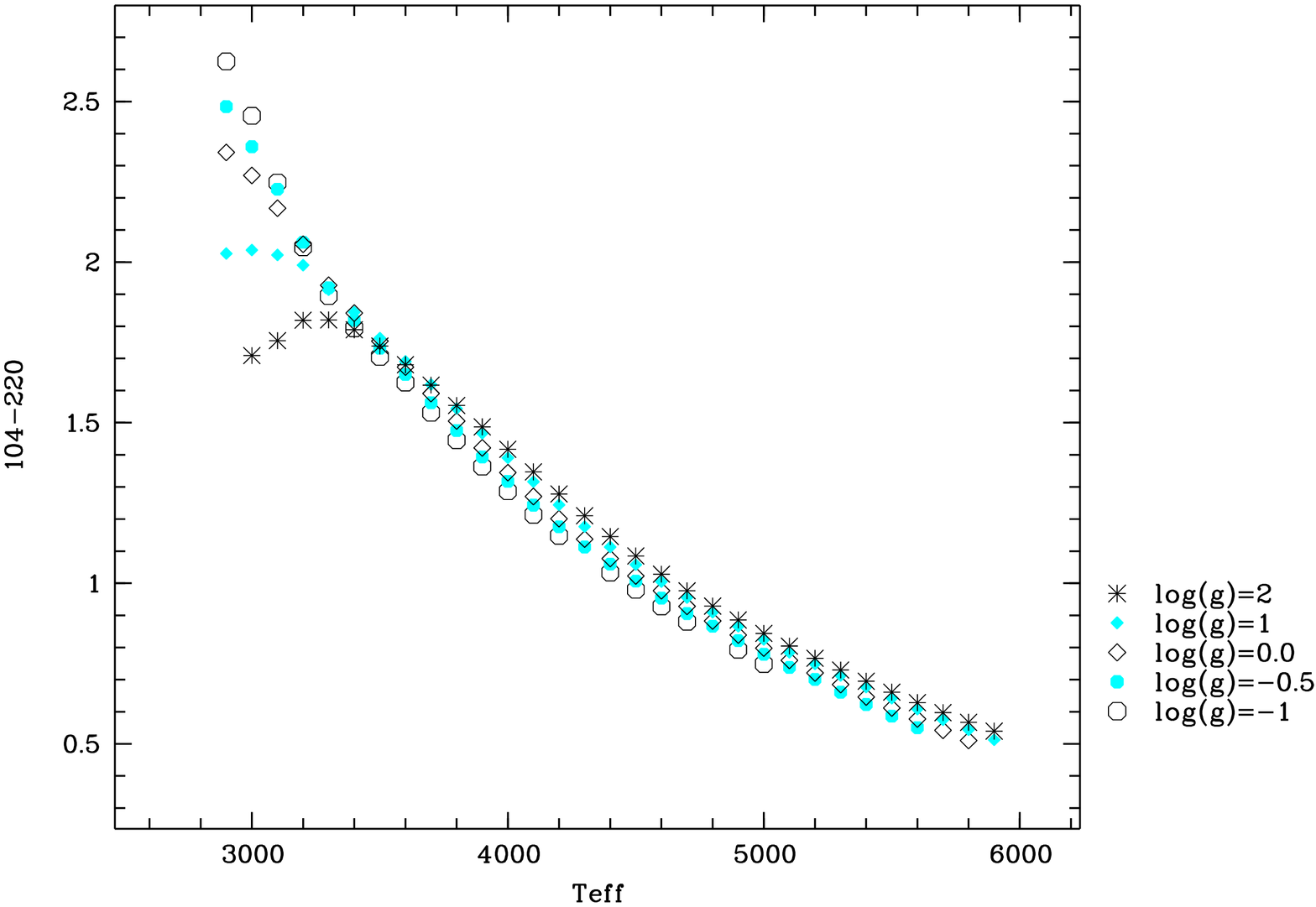}
\includegraphics[clip=,angle=270,width=0.48\textwidth]{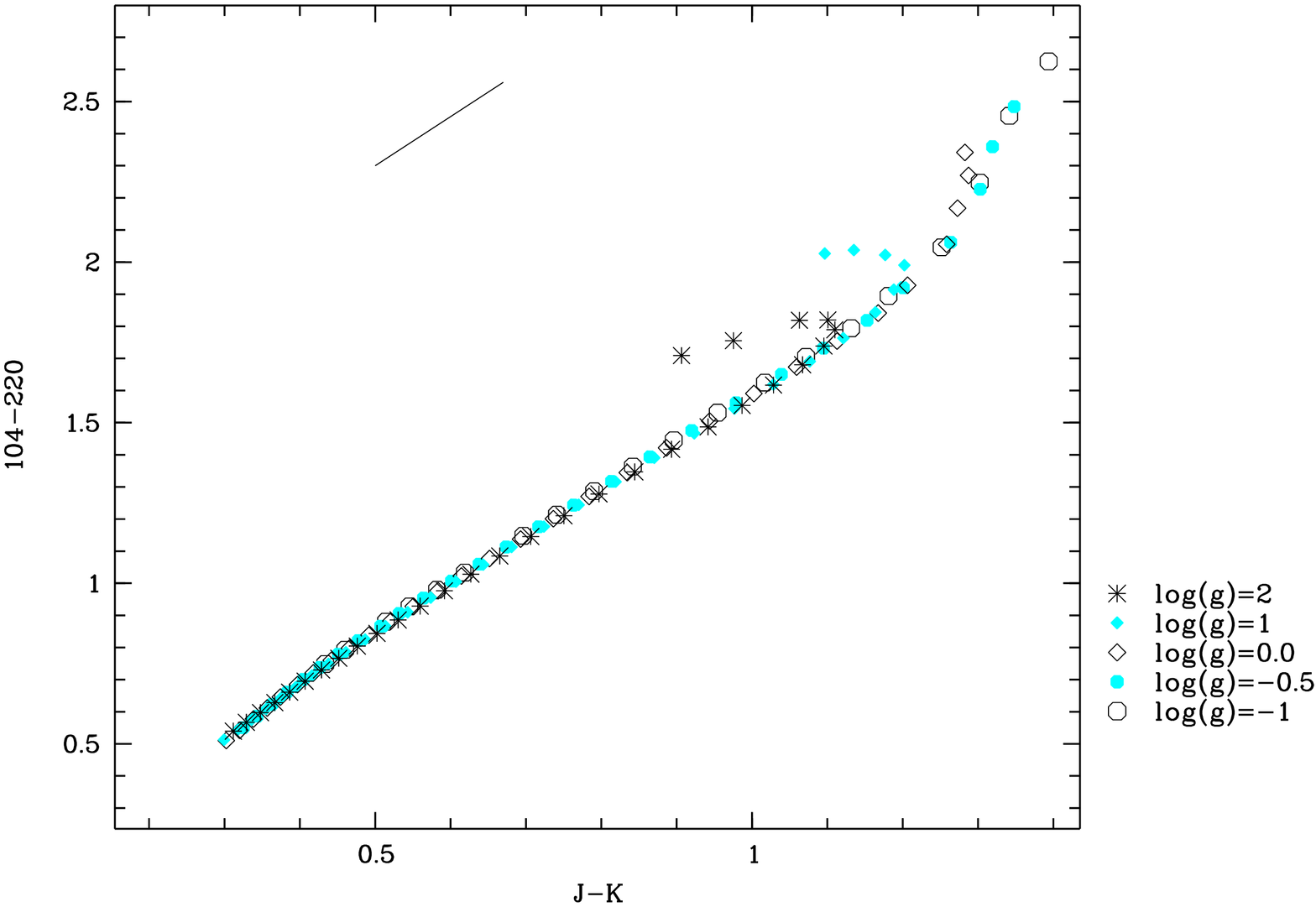}
}
\caption[]{Temperature sensitive near-IR colours 
(solar abundances, 15\,M$_{\odot}$). 
In the right panel, the effect of extinction on cool stellar spectra
is shown for A$_V$=1 (using the
extinction law of Cardelli et al. 1989, with R$_V$=3.1).}
\label{Teff_colours.fig}
\end{figure*}

\begin{figure*}
\centerline{
\includegraphics[clip=,angle=270,width=0.48\textwidth]{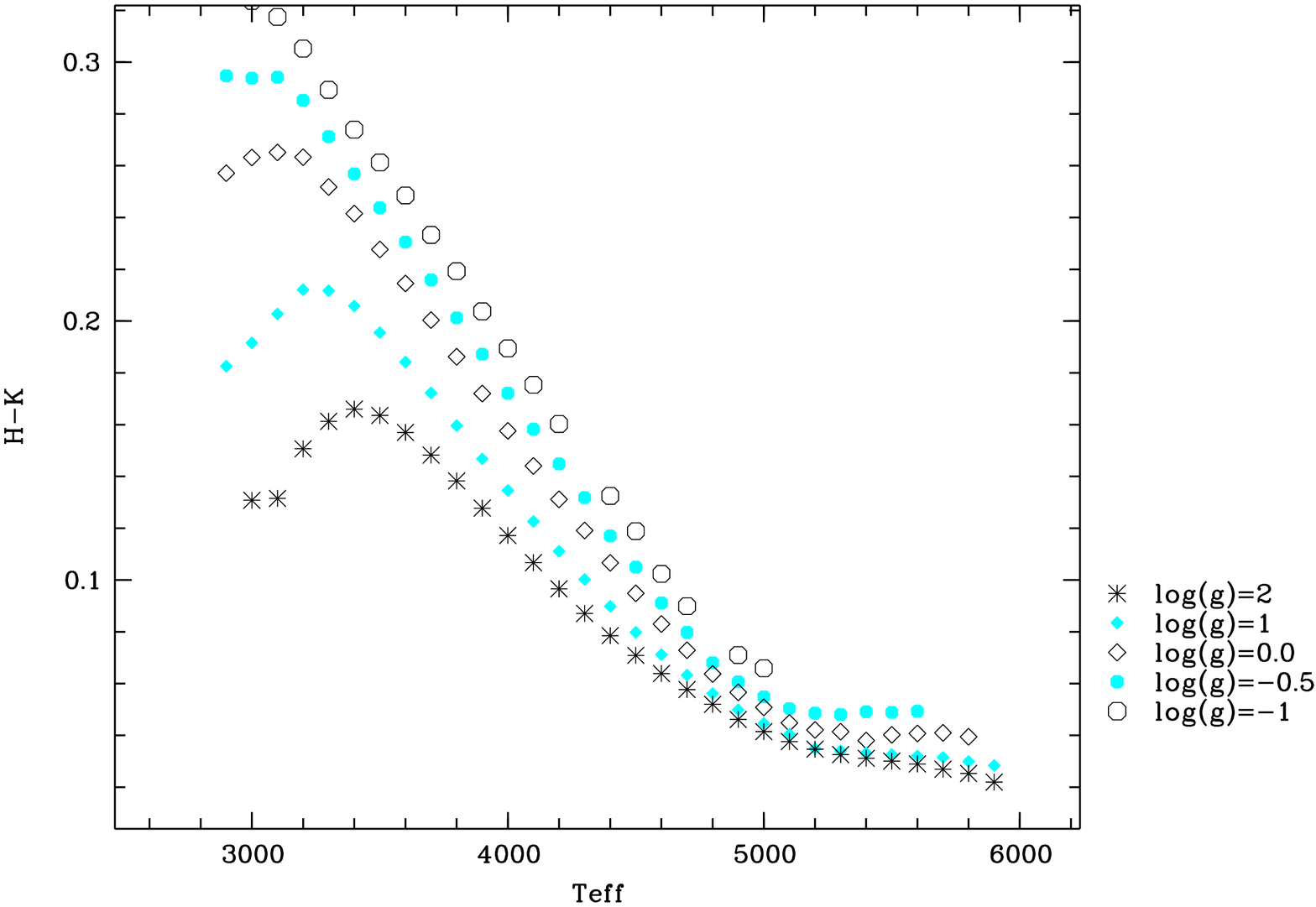}
\includegraphics[clip=,angle=270,width=0.48\textwidth]{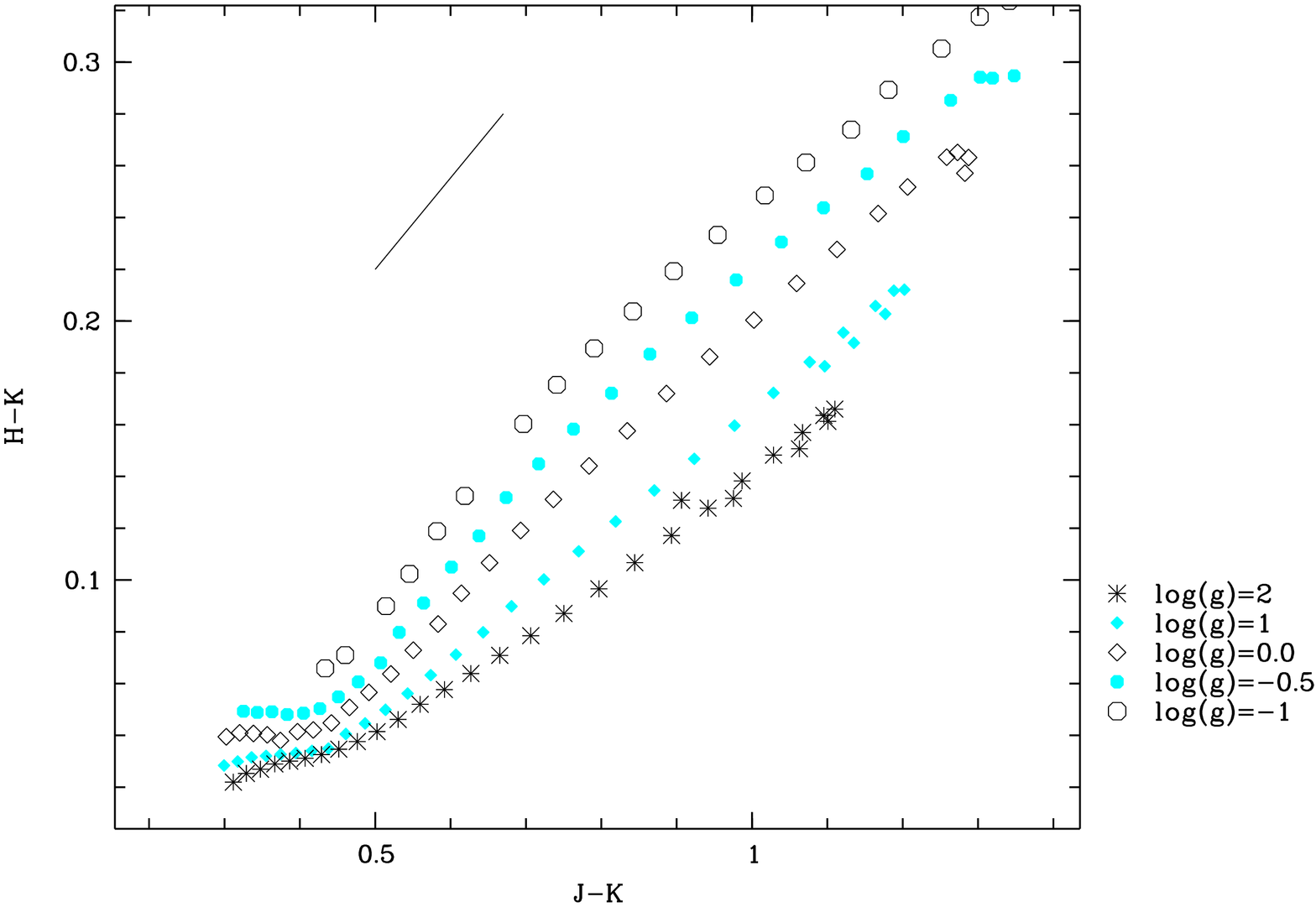}
}
\caption[]{Gravity sensitivity of H-K (solar abundances, 
15\,M$_{\odot}$).  Extinction vector as in Fig.\,\ref{Teff_colours.fig}.}
\label{logg_colours.fig}
\end{figure*}

As shown in Fig.\,\ref{Teff_colours.fig}, colours that combine flux measurements
around 1.04\,$\mu$m, in the J band and in the K band are good 
indicators of \Teff\ in theory, as their sensitivity to surface gravity is low.
Above 3400\,K, the spread in log($g$) corresponds to a full spread in 
\Teff\ of about 200\,K for the 15\,M$_{\odot}$ models (left panel). 
For 1\,M$_{\odot}$ models, the corresponding spread 
is much smaller: about 60\,K, centered on a line very close to the 
models at 15\,M$_{\odot}$ and log($g$)=0.
At the lowest temperatures, contamination of the pseudo-continuum in 
the K band with H$_2$O absorption leads to reduced fluxes in low gravity stars. 
Unfortunately, in the two-colour plots useful for observers the 
extinction vectors run almost exactly parallel to the temperature 
sequence (right panel)\,: more resolved spectral information 
is necessary to estimate an effective temperature from near-IR data.

Figure\,\ref{logg_colours.fig} illustrates the gravity dependence 
of colours involving H band fluxes.
At high gravities, the minimum of the opacity of H$^-$ around 1.6\,$\mu$m
produces a distinct hump in the H band spectra, with correspondingly
blue H-K and red J-H colours. At low gravities, molecular absorption due
mainly to CO and CN erases this continuum opacity feature.  
Such an effect was already mentioned by Bessell et al. (1991), though
their interpretation probably underestimated the r\^ole of CN as compared
to CO. The observations of Lan\c{c}on et al. (2007) and those described
in Ku\v{c}inskas et al. (2005) provide a convincing 
validation of the JHK colours of the new models.

The effect of mass on the H band flux is insignificant at log($g$)$>$0.
For lower gravities, H-K increases by up to only 0.02 magnitudes when 
going from 15 to 1\,M$_{\odot}$, at \Teff $>$4000\,K.

Switching from solar-scaled to RSG-specific abundances has the following
(small) effects on the above colours. All colours tend to become bluer.
Colour differences in H-K and J-H remain smaller than 0.04\,mag (and 
are $<$0.02\,mag for most stellar parameters). The colour index 104$-$220 
(defined in Tab.\,\ref{indexdef.tab}) is reduced by up to 0.08\,mag. 
The bolometric corrections to the K band,
BC(K), are essentially unchanged (the \Phoenix\ values agree with
those of Levesque et al. 2005 to within a few percent between
3400 and 4300\,K). Effects this small would be difficult 
to exploit on real stellar spectra.

\subsection{Molecular indices}

\subsubsection{CO}

CO is a long known indicator of luminosity class (Baldwin et al. 1973). It is 
sensitive to gravity and effective temperature, but also to
metallicity and micro-turbulence.  As indicated previously, 
a constant micro-turbulent velocity of 2\,km.s$^{-1}$
is used in this paper except in a few models.
Large micro-turbulent velocities deepen the 1st overtone band of CO more
than the 2nd overtone band, because line saturation is more important in
the former than in the latter (e.g. Origlia et al. 1993, 1997 
and refs. therein).  

\begin{figure*}
\centerline{
\includegraphics[clip=,angle=270,width=0.55\textwidth]{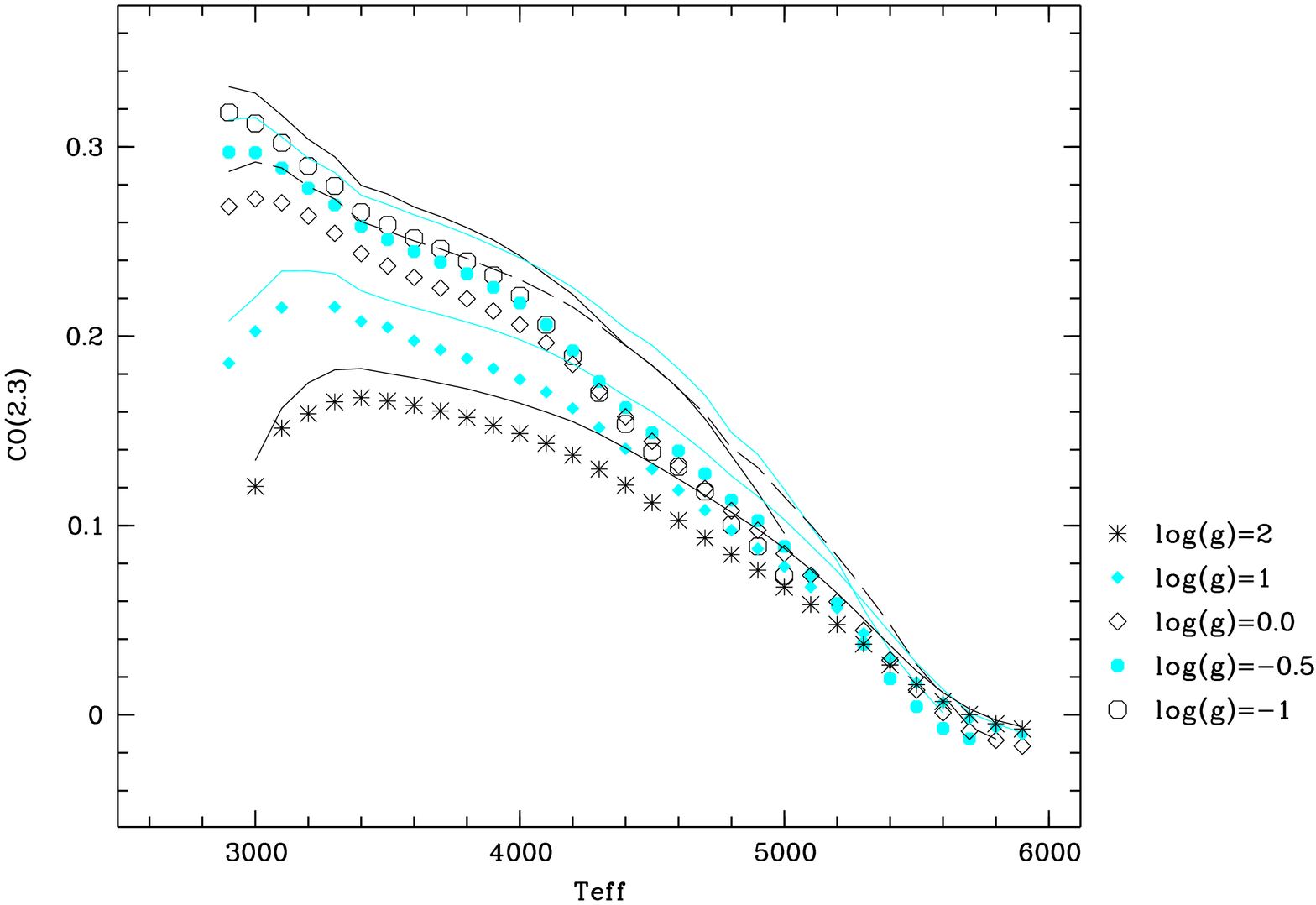}
\hspace{-2.0cm}
\includegraphics[clip=,angle=270,width=0.55\textwidth]{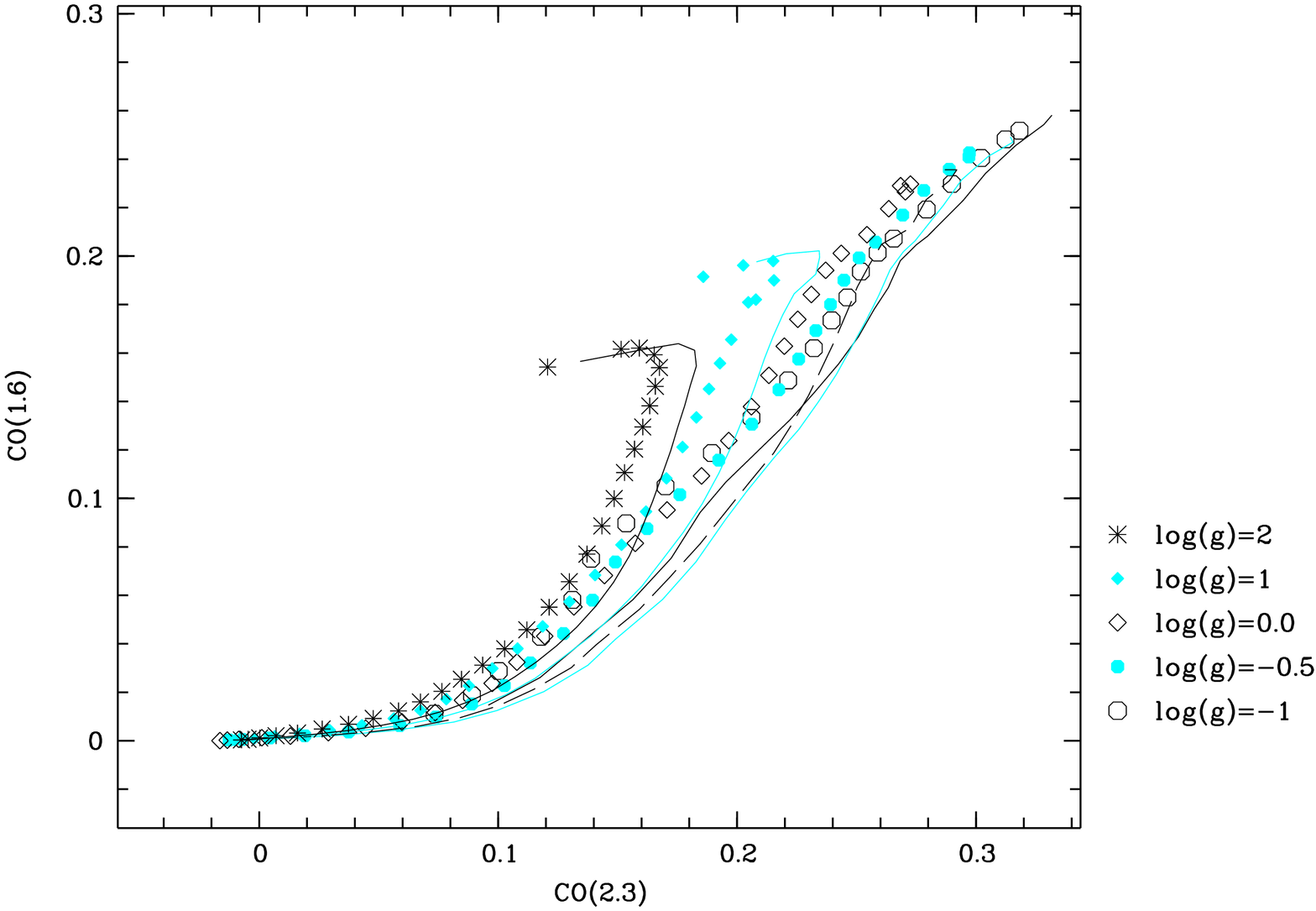}
}
\caption[]{Measurements of the strength of the 1st overtone CO 
band at 2.29\,$\mu$m and of the 2nd overtone CO band at 1.62\,$\mu$m
in the model spectra (15\,M$_{\odot}$).
Symbols: temperature sequences at the indicated gravities, 
for RSG-specific abundances. Lines: corresponding sequences
for solar-scaled abundances 
(black and light-coloured lines alternate, and 
the dashed line has log($g$)=0). 
Note that at a given CO\,(2.3), xCO\,(1.6) 
tends to be weaker in low gravity stars.
}
\label{Teff_CO.fig}
\end{figure*}

On the left panel of Fig.\,\ref{Teff_CO.fig}, the changes of the
1st overtone CO band at 2.29\,$\mu$m with gravity, temperature and 
surface abundances are 
shown. As commonly found, CO increases with decreasing temperatures
and gravities. The CO strength progressively levels off when log($g$) 
takes negative values (i.e. the further dependence on $g$ is negligible). 
Contamination by H$_2$O at high log($g$) 
produces a drop of the CO index below 3200\,K. Switching from solar
to RSG-specific abundances reduces the CO strength generally by small amounts.
The effect is maximum around 4500\,K in low gravity models\,: RSG-specific
models with log($g$)=$-1$ at 4500\,K 
have the same CO index as solar abundance models
with log($g$)=2, or alternatively with log($g$)=$-1$ but 4800\,K.

The effects of log($g$), \Teff\ and abundances on the apparent strength of
the 2nd overtone CO band at 1.62\,$\mu$m 
are similar, with two notable exceptions.
First, the effects of changes in the abundance ratios are smaller than
at 2.29\,$\mu$m. Second, low gravity saturation is reached earlier.  The 
result is summarized in the right panel of Fig.\,\ref{Teff_CO.fig}. 
In particular, {\em low gravity stars tend to have weaker CO bands around
1.6\,$\mu$m than high gravity ones, at a given strength of the 
2.29\,$\mu$m band.} 
Contamination of the H-band fluxes by CN absorption contributes to
producing this trend, as already hinted at by Wing \& Spinrad 1970.

Moving down from 15\,M$_{\odot}$ to 1\,M$_{\odot}$ has a negligible effect 
on the near-IR CO bands for log($g$)$>$0. At lower gravities, the 1\,M$_{\odot}$
CO bands are weaker than the 15\,M$_{\odot}$ bands (the effect is stronger
for the 2.3\,$\mu$m band than for the 1.6\,$\mu$m bands).

\subsubsection{CN}

\begin{figure*}
\centerline{
\includegraphics[clip=,angle=270,width=0.55\textwidth]{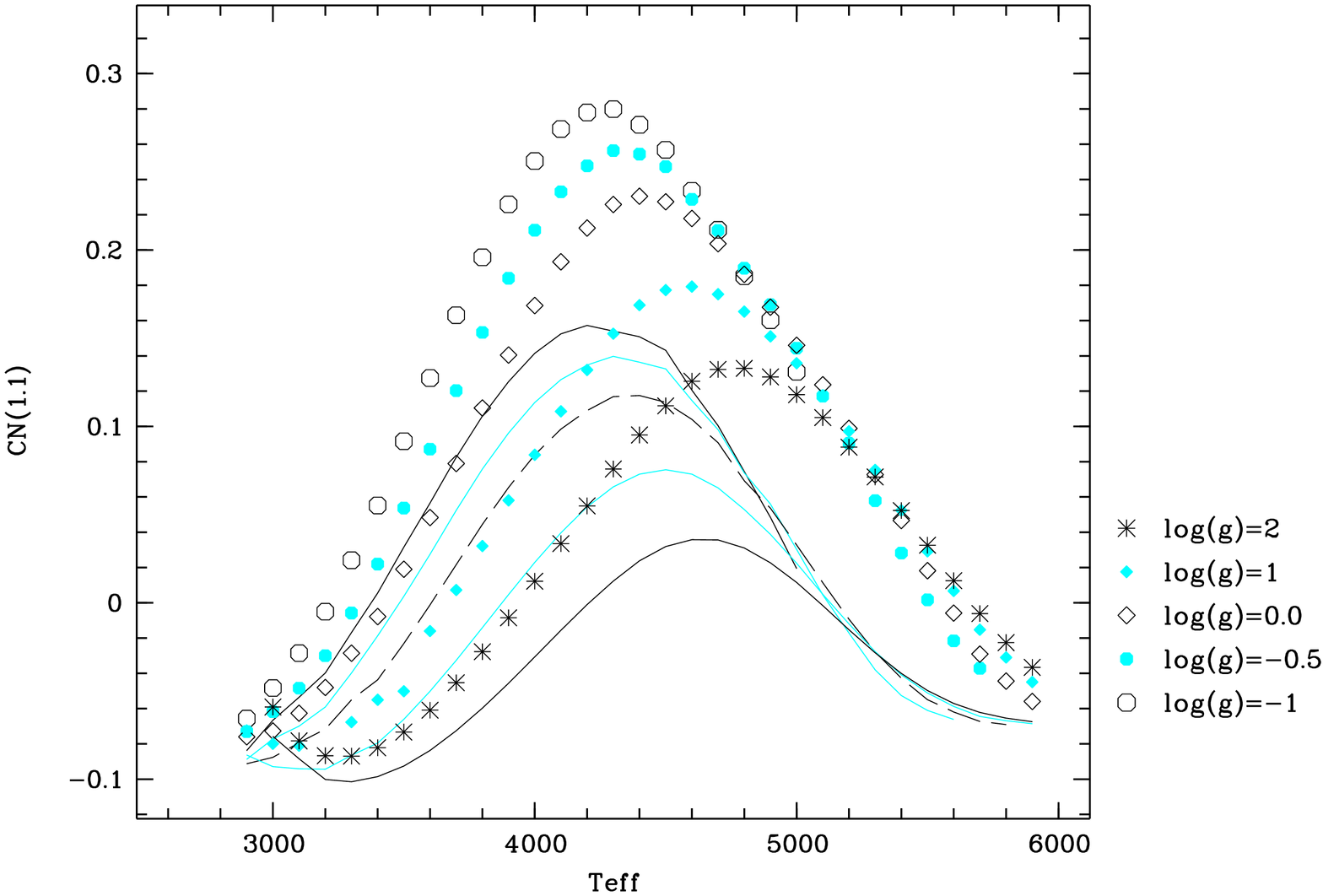}
\hspace{-2.0cm}
\includegraphics[clip=,angle=270,width=0.55\textwidth]{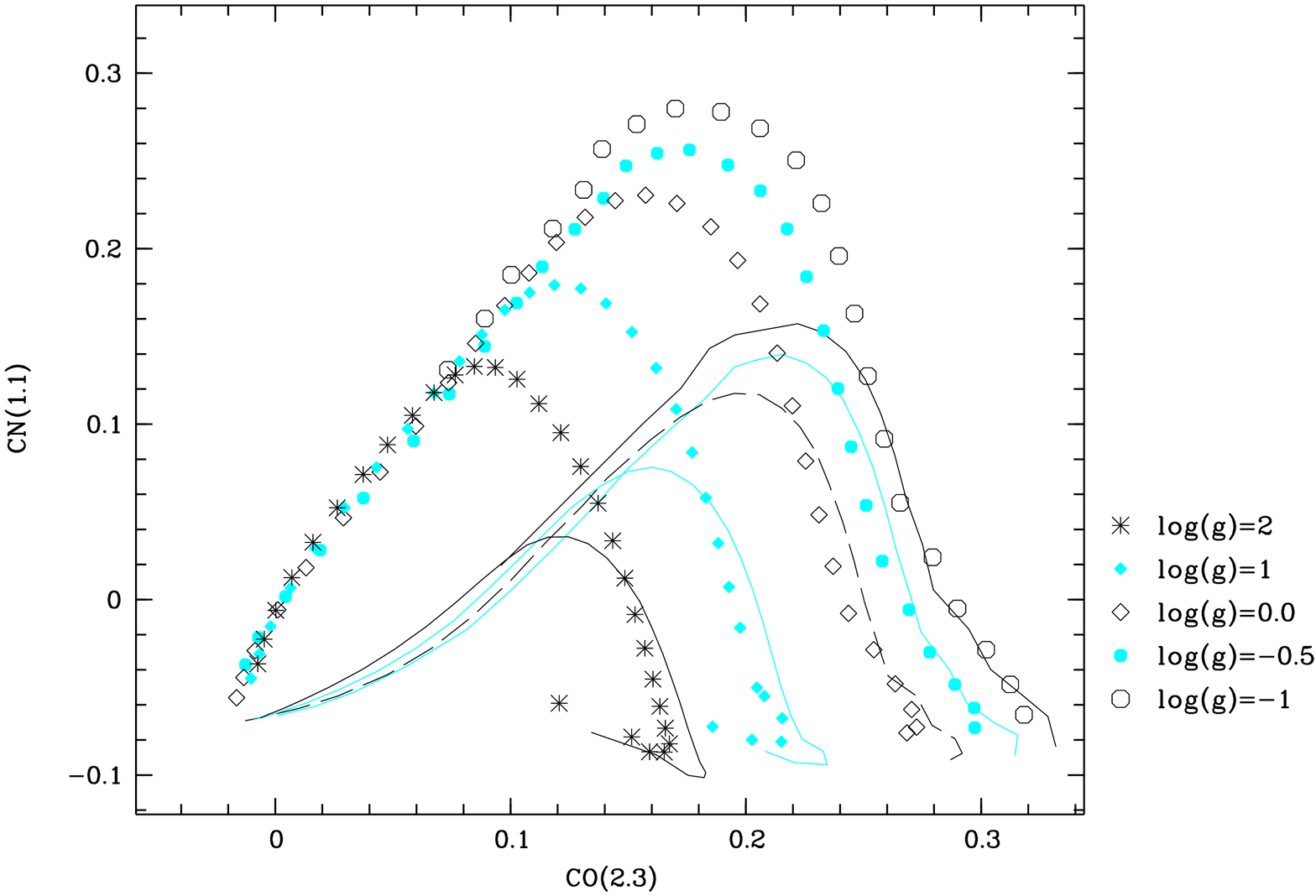}
}
\caption[]{Measurements of the strength of the CN band at 1.1\,$\mu$m in
the model spectra (15\,M$_{\odot}$).
Symbols and lines are as in Fig.\,\ref{Teff_CO.fig}. The main effect of
a decrease of mass (15$\rightarrow$1\,M$_{\odot}$) 
is a shift of the models at log($g$)$\leq$0
to the left by up to 0.04\,mag in the right hand panel.
}
\label{Teff_CN.fig}
\end{figure*}

CN displays prominent near-IR absorption bands that have been studied
extensively in the context of carbon star models (e.g. Loidl et al. 2001).
The CN bands are prominent in red supergiants as well (White \& Wing, 1978). 
While in carbon stars essentially all surface oxygen is locked into CO,
CN coexists with other oxides in the atmospheres of red supergiants.  

The behaviour of CN bands with varying model parameters is complex, as shown in
Fig.\,\ref{Teff_CN.fig}. Bessell et al. (1989) describe the decrease
of the CN 1.1\,$\mu$m band strength 
with decreasing effective temperature below 3800\,K,
as well as its gravity dependence (stronger CN for lower gravities).
Our more extended temperature range shows that the maximum CN strength
is reached between 4200 and 4800\,K. Both the location of the maximum
and its actual strength depend on surface gravity, and on the chemical
composition of the atmosphere. {\em CN bands are strongly enhanced
in models with RSG-specific abundance ratios.} 
The effect of mass is small. 
CN is also enhanced when larger
micro-turbulent velocities are assumed (Tsuji 1976). In empirical samples,
spectra with strong CN absorption bands compared to their CO bands are 
candidates for modified surface abundances.

\subsubsection{Other molecular bands longwards of 1\,$\mu$m.}

\begin{description}

\item[H$_2$O.] H$_2$O appears very abruptly in the models below
a gravity-dependent threshold temperature: 3600\,K at log($g$)=1,
3100\,K at log($g$)=$-1$ (based on measurements in the K-band wings of 
the H$_2$O band centered near 1.9\,$\mu$m). Below this threshold \Teff, 
higher gravities lead to stronger spectral bands. Varying the mass between
1 and 15\,M$_{\odot}$, or switching to RSG-specific
abundances, has only small effects on the H$_2$O bands. 

\item[TiO.] Near-IR TiO bands around 1\,$\mu$m (TiO $\delta$, $\Delta\nu=-1$)
and 1.25\,$\mu$m (TiO $\phi$, $\Delta \nu = -1$) 
appear progressively in the models below a gravity-dependent temperature:
$\sim$3600\,K at log($g$)=1, $\sim$3400\,K at log($g$)=$-$2 
(based on visual inspection of the spectra in that region). 
Other near-IR TiO bands longwards of 1\,$\mu$m,
such as the $\phi, \Delta \nu=0$ band near
1.12\,$\mu$m are hidden in CN absorption.
Again, varying the mass between
1 and 15\,M$_{\odot}$, or switching to RSG-specific
abundances, has only small effects.  
We note that the next version of \Phoenix\ calculations will include an
update of the TiO partition function and of the electron $f$-values of the 
TiO bands for the AMES TiO line list (Schwenke 1998), which appears to
improve spectral synthesis results for M dwarfs (Allard et al., in preparation).

\item[VO.] The 1.05\,$\mu$m VO band (VO A$-$X $\Delta v=0$)
is significant in the model spectra 
only at \Teff$\leq$3200\,K for log($g$)$\leq$1. 
The effect of mass or abundances is small.

\end{description}


\section{Models versus data in two-colour plots}                   
\label{twocolour.sec}

\begin{figure*}
\centerline{
\includegraphics[clip=,angle=270,width=0.48\textwidth]{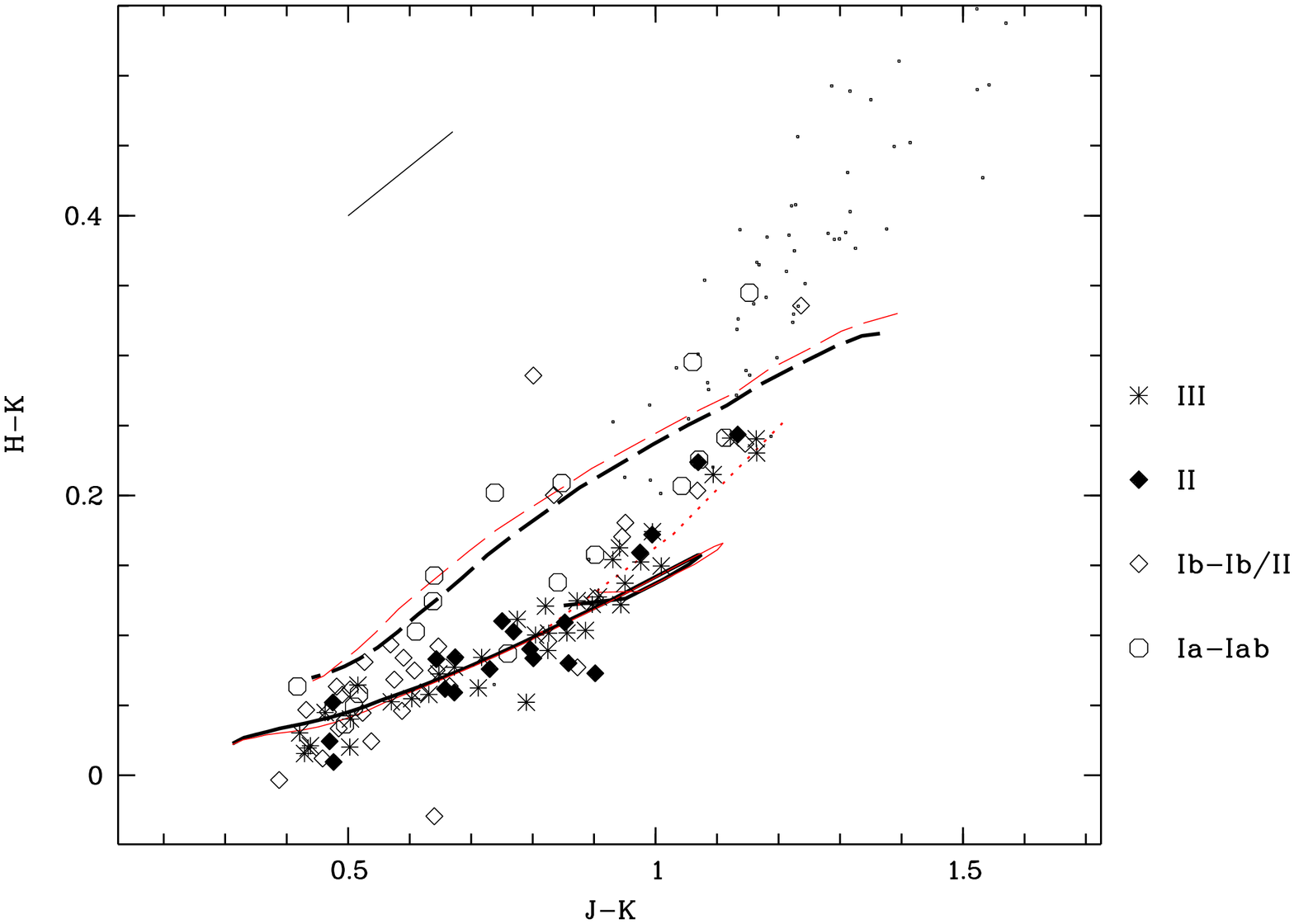}
\includegraphics[clip=,angle=270,width=0.48\textwidth]{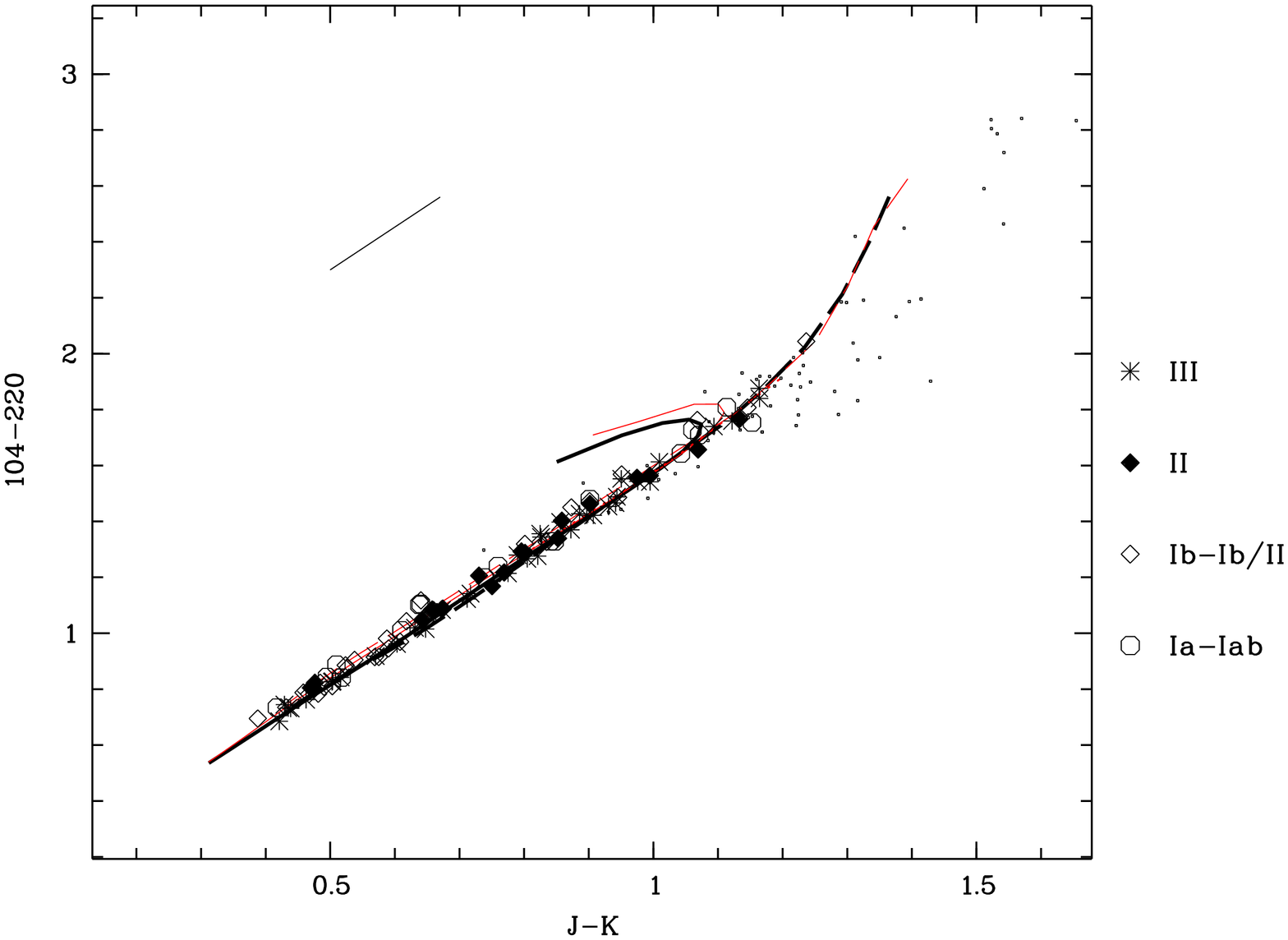}
}
\caption[]{Two-colour plots with observational and calculated data. 
The thin lines are \Teff\ sequences
for solar abundances, the thick lines are for RSG-specific abundances.
Solid lines are at log($g$)=2, dashed lines at log($g$)=$-1$. The dotted
line follows models along an illustrative red giant branch at solar
metallicity ([\Teff (K),log($g$)]=[4200,2], [3800,1], [3400,0]). Symbols 
as explained are measurements on {\em dereddened} versions of 
the spectra of Lan\c{c}on et al. 2007 (see Sect.\,\ref{spectra.sec}),
and dots O-rich Miras from Lan\c{c}on \& Wood 2000 (not dereddened). 
The reddening vectors are as in Fig.\,\ref{Teff_colours.fig}.
%
}
\label{data_colours.fig}
\end{figure*}

In this and the following sections, we compare the models with 
the data collected by Lan\c{c}on \& Wood (2000) and Lan\c{c}on et al. (2007).
Both sets provide spectra at a spectral resolving power 
of order 1000 between 0.97\,$\mu$m (sometimes 0.81\,$\mu$m) and 2.4\,$\mu$m.
The first set adds low resolution extensions through the optical
range down to 5100\,\AA\ for a few of the stars.
The merged sample contains luminous stars of spectral types
G3I to M5I, G3II to M3II, and G5III to M5III, as well as asymptotic giant
branch variables for comparison.

As shown in Fig.\,\ref{data_colours.fig}, the agreement between models and
data is excellent in near-IR two-colour plots
once extinction has been accounted for (see Sect.\,\ref{spectra.sec}).
Note that in this and the following figures the lines join models 
at constant log($g$), while the data should follow real red giant or
red supergiant branches. Evolutionary tracks predict that the 
warmer giants have log($g$)$\geq$2 while the cool ones reach 
log($g$)$\simeq$0. Red supergiants of various luminosity classes
are expected to have log($g$)$\leq$0.5. 

\begin{figure*}
\centerline{
\hspace*{-0.06\textwidth}
\includegraphics[clip=,angle=270,width=0.4\textwidth]{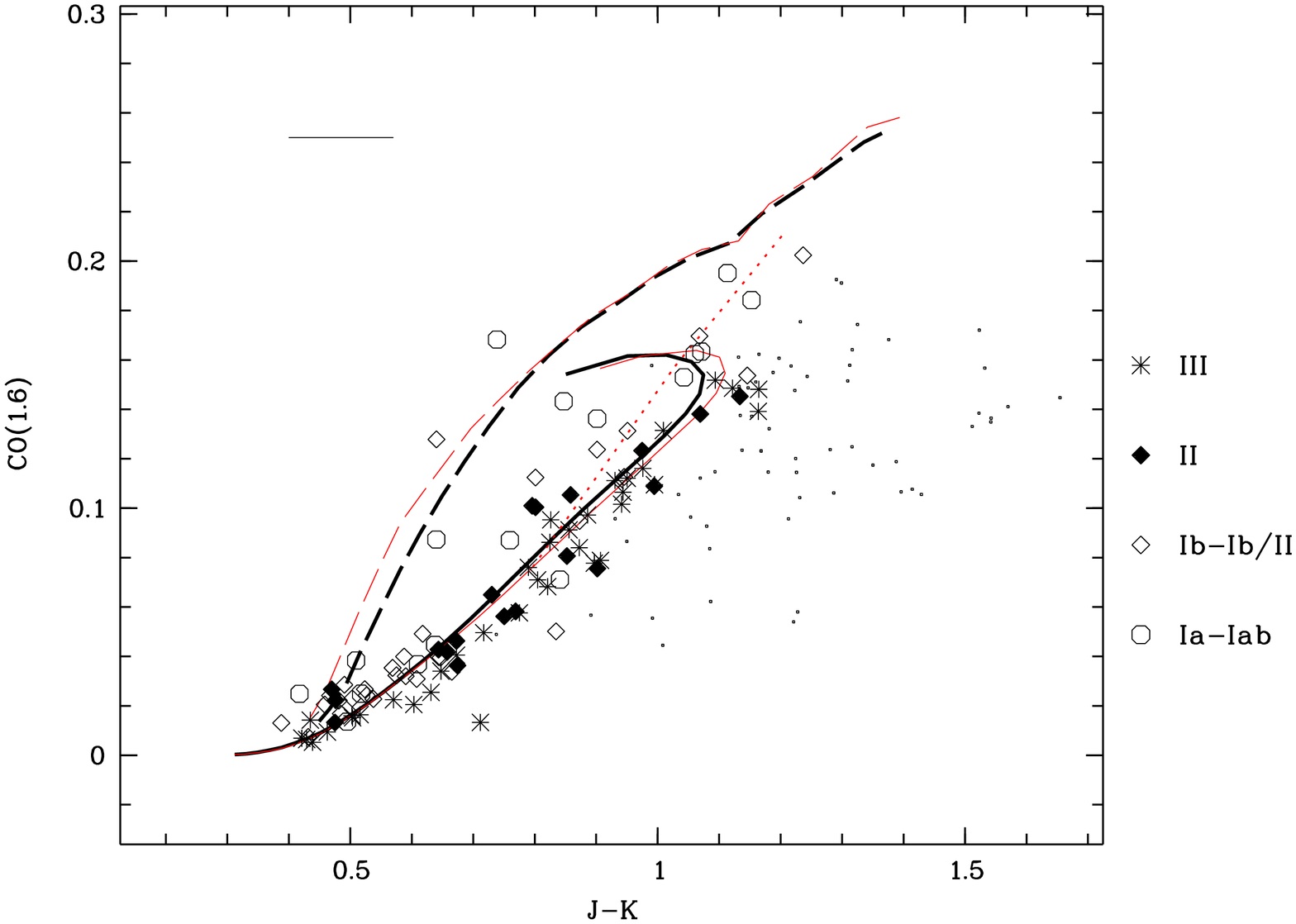}
\hspace{-0.082\textwidth}
\includegraphics[clip=,angle=270,width=0.4\textwidth]{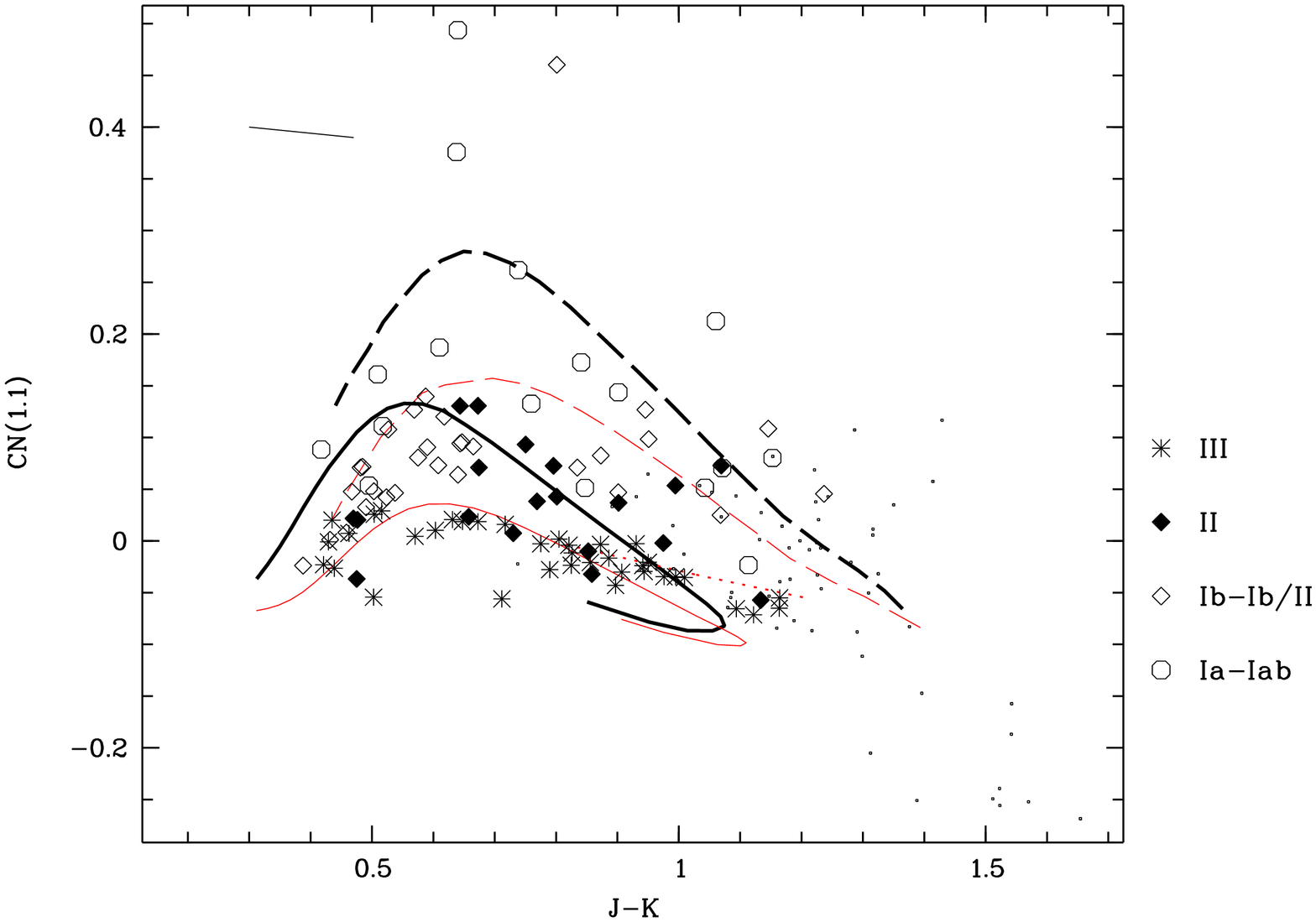}
\hspace{-0.082\textwidth}
\includegraphics[clip=,angle=270,width=0.4\textwidth]{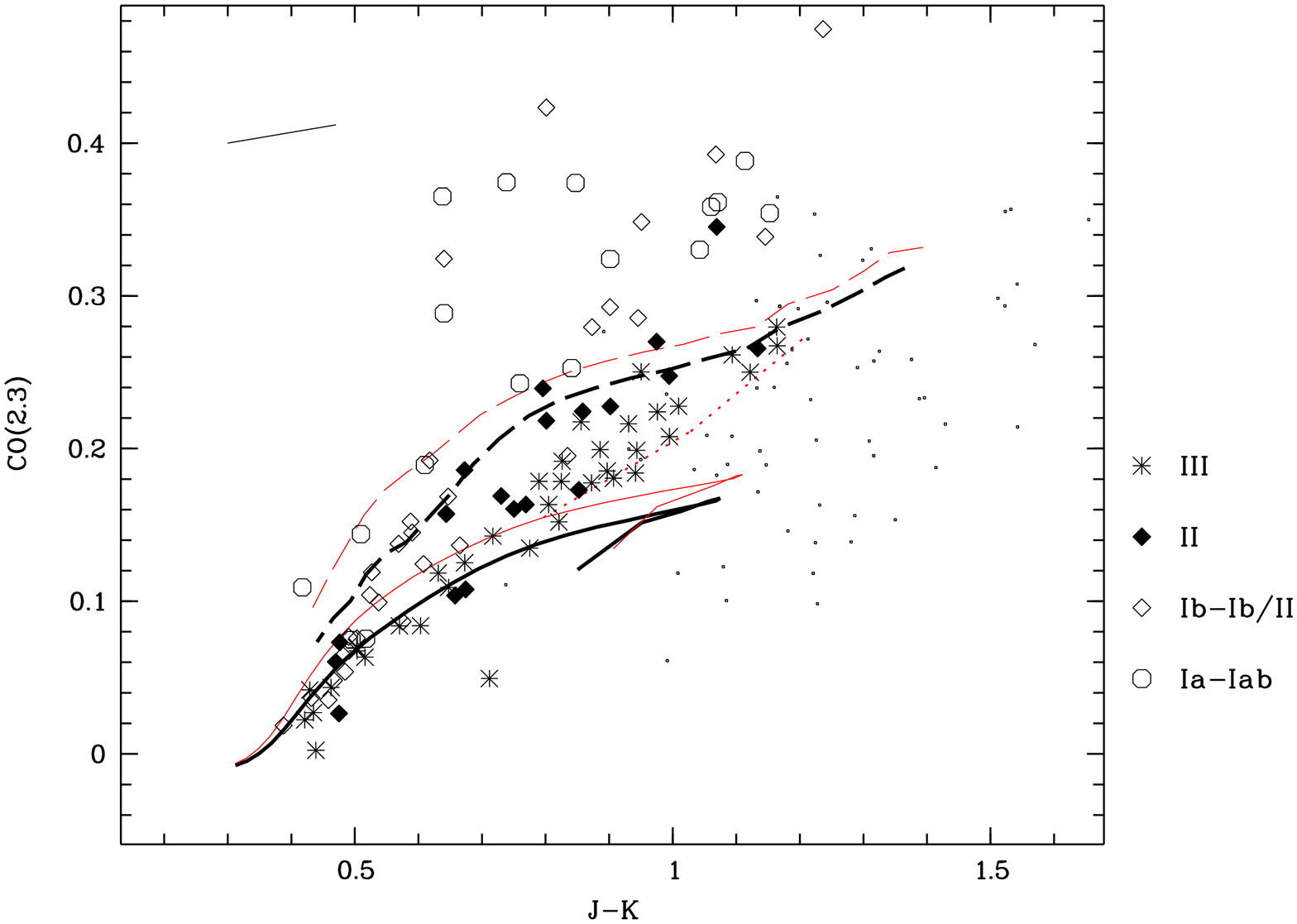}
}
\caption[]{Plots of molecular indices vs. colours with dereddened data.
Symbols and lines
are as in Fig.\,\ref{data_colours.fig}.
}
\label{data_indices_colours.fig}
\end{figure*}

\begin{figure*}
\centerline{
\hspace*{-0.06\textwidth}
\includegraphics[clip=,angle=270,width=0.4\textwidth]{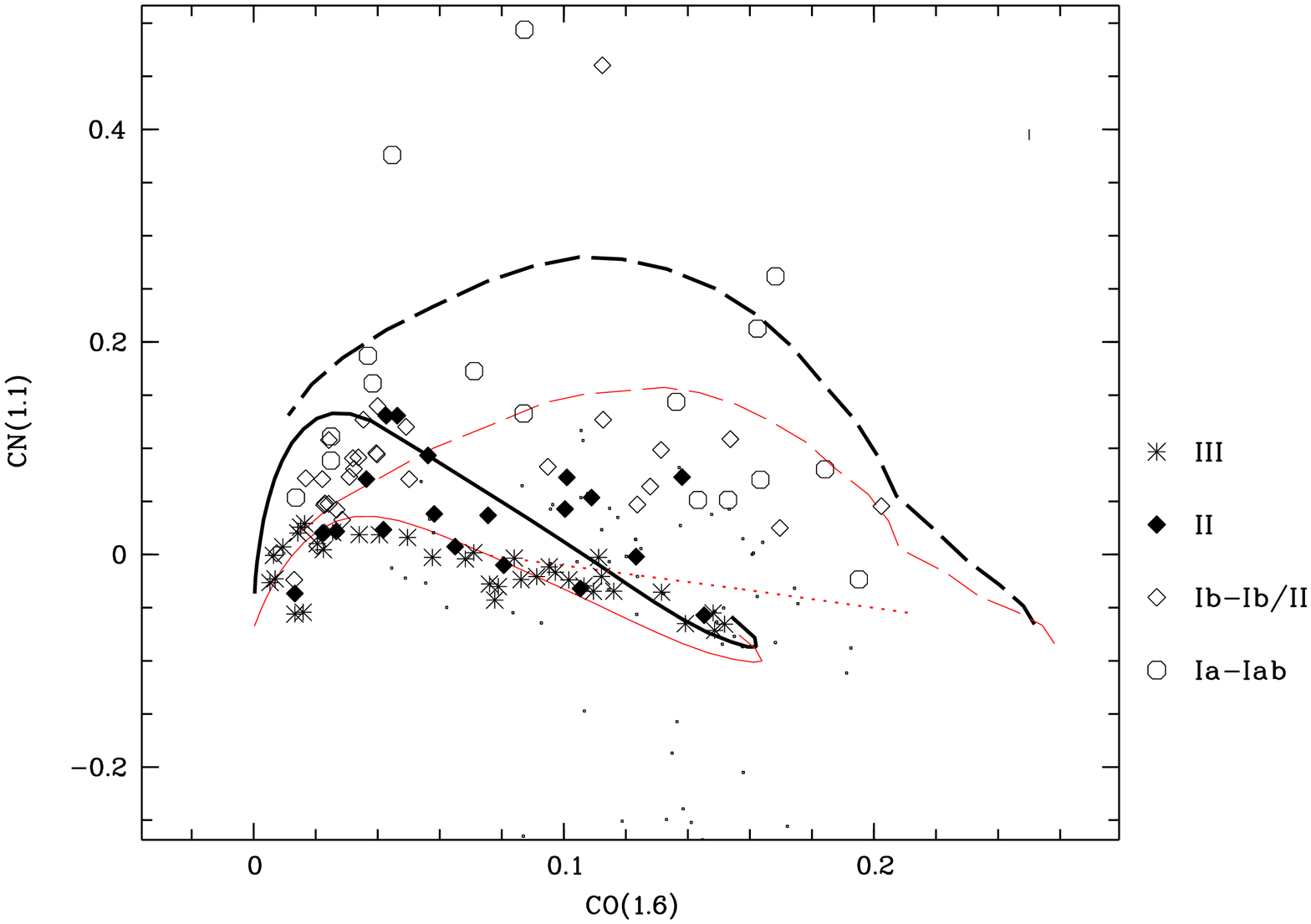}
\hspace{-0.082\textwidth}
\includegraphics[clip=,angle=270,width=0.4\textwidth]{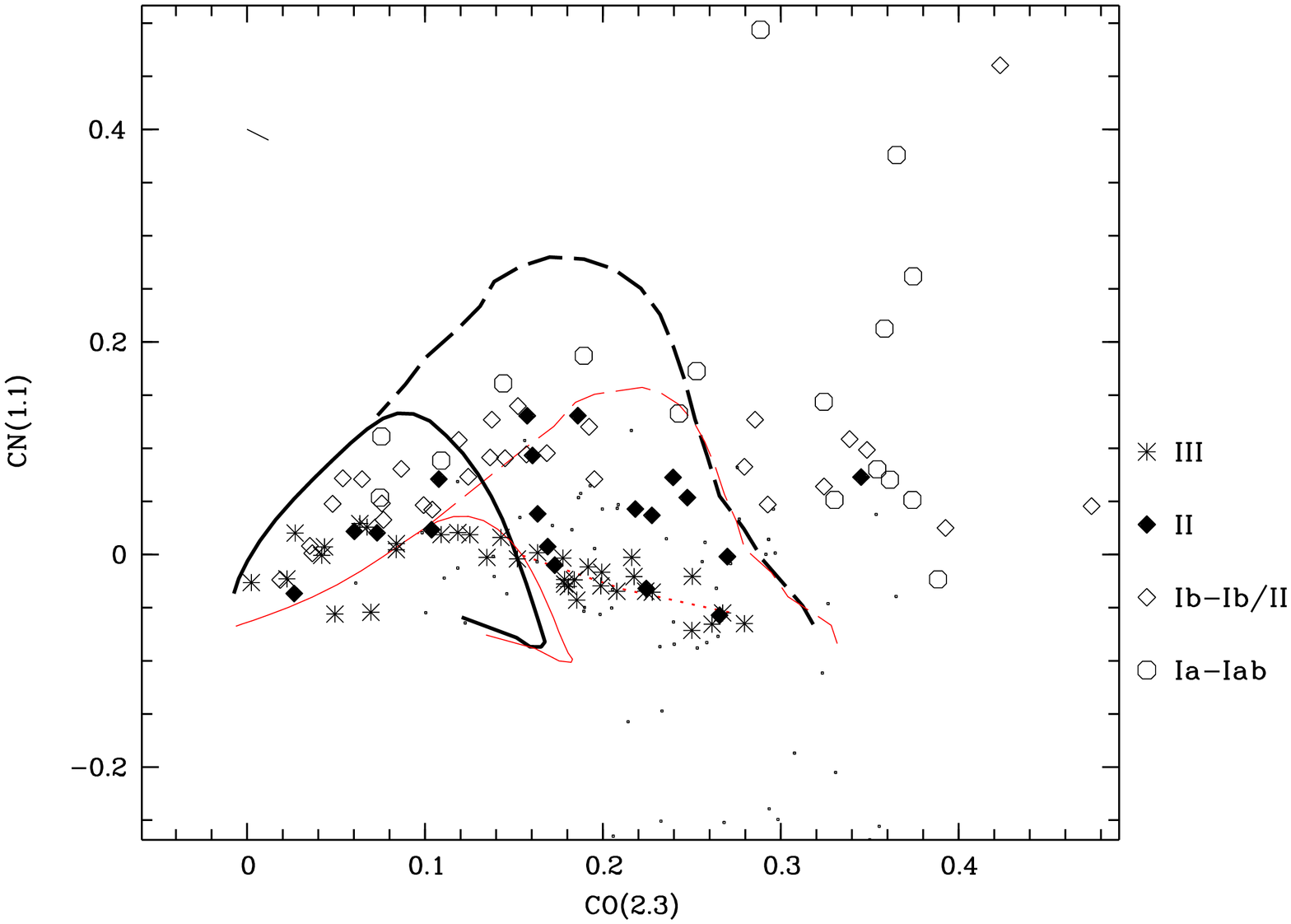}
\hspace{-0.082\textwidth}
\includegraphics[clip=,angle=270,width=0.4\textwidth]{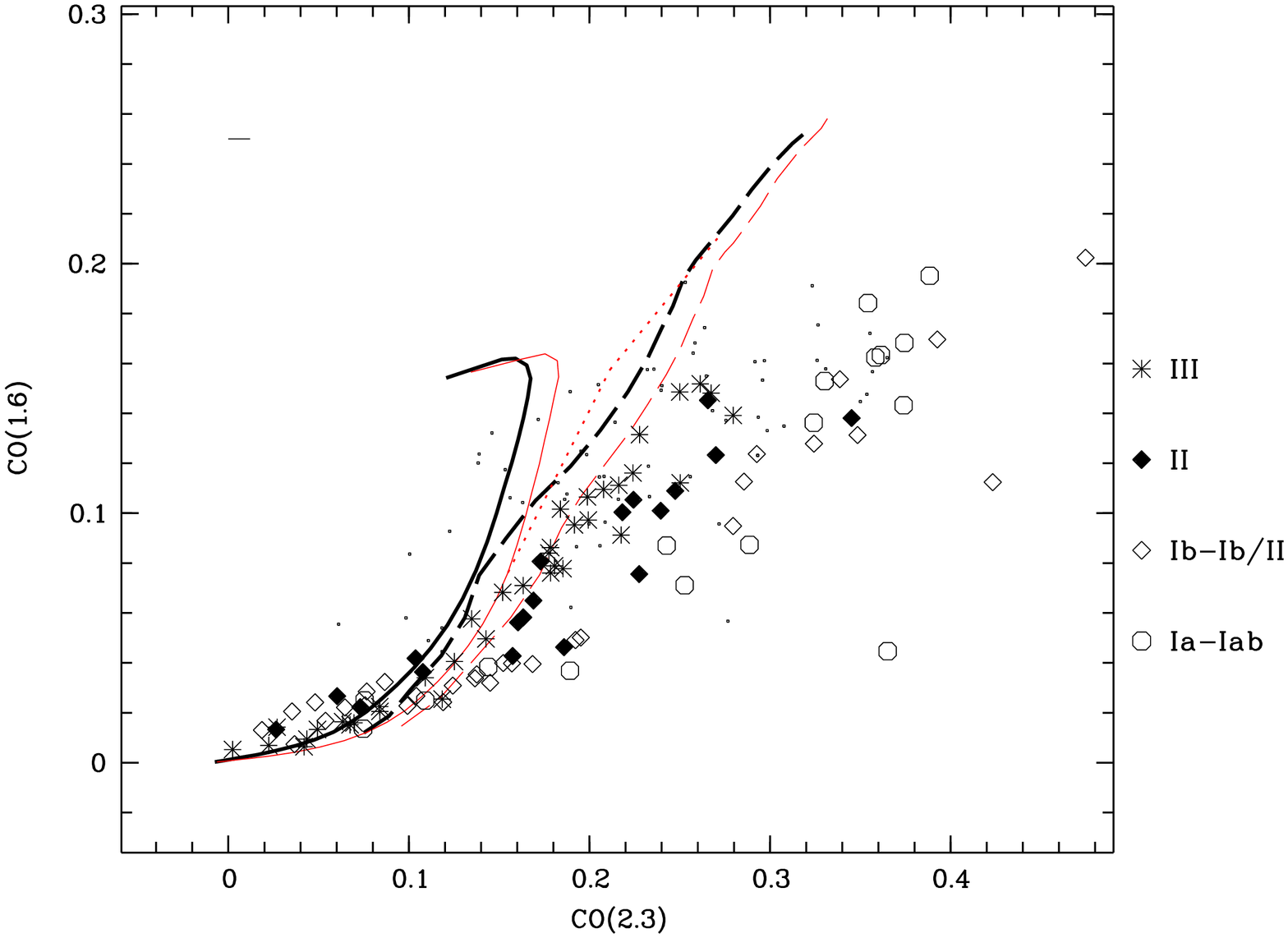}
}
\caption[]{Molecular index plots with dereddened data. Symbols and lines
are as in Fig.\,\ref{data_colours.fig}.
}
\label{data_indices.fig}
\end{figure*}

Figure\,\ref{data_indices_colours.fig} combines measurements 
of the first and second overtone CO bands and the 1.1\,$\mu$m CN band
with J-K.  Agreement between solar metallicity models and empirical 
data is good for {\em giant} stars. 
The strongest offset is noted for the second overtone CO
bands in the H window, which tend to be too strong in
the cool giant star models. The figures also suggest that 
modeled first overtone CO bands might be slightly too weak at 
low $\Teff$. Because extinction affects J-K, 
two-index diagrams with a negligible sensitivity to reddening are
presented in Fig.\,\ref{data_indices.fig}. The same conclusions hold.
The CO line list data are from
Goorvitch \& Chackerian (1994\,a,b), and are known to
work very well in the case of M dwarfs. Therefore, it is unlikely
that the line list data for CO is the cause of the CO band
discrepancies.
We note that $^{13}$CO contributes to the measured strength of the 
first overtone CO bands, and that changes in the $^{13}$C abundances 
induced by stellar evolution may be responsible for some systematic effects.
Slightly larger micro-turbulent velocities could improve the ratio of the 
first to the second overtone CO band strengths, but would also affect
the CN bands. 
The outlier giant star near J-K=0.7 with weak band strengths is 
the only Population II star of the observed sample (HD\,218732). Eye
inspection of its spectrum reveals a metal poor atmosphere immediately. 
By contrast, the other giant stars appear to form a reasonably 
homogeneous sample in terms of metallicity. 

While the trends with gravity present in the predicted molecular bands
agree qualitatively with those observed, the molecular bands of only
a fraction of the observed red {\em supergiants} can be reproduced
quantitatively. Models with RSG-specific abundances are favoured 
for a significant number of these objects, which show stronger CN
bands than the solar metallicity models can produce. However, the CN and 
CO measurements show that despite the improvement achieved with the 
adopted changes in surface abundances,
the model parameters explored here are not 
able to account for the whole range of molecular band strengths observed.
Models with larger micro-turbulent velocities reach further into
some of the areas occupied by real red supergiants, 
justifying the ongoing extension
of the model grid. Alternatively, some red supergiants in 
nature may require even higher $^{14}$N abundances than tested here 
or effective gravities lower than log($g$)=$-1$.

Comparisons between models and data on the star-by-star
basis are provided in the following section.

\section{Direct comparison between observed and theoretical spectra}                   
\label{spectra.sec}

\subsection{Method}
\label{method.sec}

The comparison between models and data is based on the computation
of reduced $\chi^2$ differences. The theoretical spectra are 
smoothed to the resolution of the data, using gaussian broadening 
functions with adequate widths (note that for the optical spectra, whose
resolution was seeing-dependent, we adopt a single smoothing value
which in some cases underestimates the actual $\Delta\lambda$).
They are then resampled to the wavelength step of the data, i.e. 5\,\AA.
A window function is used to eliminate the spectral intervals most
strongly affected by telluric absorption, around 1.15, 1.4 and 1.9\,$\mu$m.
 
The rms noise of the data is modelled as being proportional to the 
square root of the signal. Numerical values given below 
assume an average signal-to-noise ratio of 50. This simple noise model is 
a reasonable representation of the typical high frequency noise of the 
data. The additional uncertainties due to flux calibration errors are not
explicitly accounted for. They lead mainly to uncertainties in the estimated
extinction values.  A further discussion of the effects of the
weighting of the data is provided in Sect.\,\ref{discussion.mass.sec}.

A mass of 1\,M$_{\odot}$ is assumed for giants and bright giants, a
mass of 15\,M$_{\odot}$ for supergiants (see Sect.\,\ref{discussion.sec}).
For each empirical spectrum, the adopted algorithm loops 
through all model temperatures
and gravities (separately for the two sets of abundances). 
At each of these points, it also loops through an adequate
set of extinctions (using the extinction law of Cardelli et al. 1989
with $R_V=3.1$), and minimizes the $\chi^2$ with respect to the extinction
parameter $A_V$. The step in $A_V$ is of 0.05\,mag. 
A $\chi^2$ map is produced in \Teff$-$log($g$) space,
and the 9 best fits (of the \Teff$-$log($g$)-A$_V$ space) are plotted
for inspection. The $\chi^2$ value of the 9th best fit is typically
higher than for the 1st best fit by 10\,\% when the best fits are good, and 
by only a few percent when they are poor.  Typical uncertainties on
the derived parameters in the case of {\em good} fits 
are $\pm$\,100\,K in \Teff, and $\pm$\,0.2 in A$_V$. 
For these good fits, gravity is usually determined to better than one step 
within our set of models (log($g$)=$-1,-0.5,0,1,2$).
Preliminary models with micro-turbulent velocities 
larger than 2\,km\,s$^{-1}$ were 
tested only for supergiant star spectra and, in this paper, 
they are only discussed in cases where the initial fits were poor. 

The method is robust with respect to errors in the positions of the individual
molecular lines that jointly define the pseudo-continuum of the 
spectra at the resolution of interest here. Such errors were noted as
a difficulty in the measurement of individual metal lines by Vanhollebeke
et al. 2006. In order to verify this, we added random noise to 
the data at the level of a few percent, i.e. enough to completely alter the 
apparent positions of the blended CN lines between 2 and 2.3\,$\mu$m.
Differences in the derived parameters were well within the 
uncertainties stated above.

\subsection{Giant stars}

The data sample contains 35 stars of class III with spectral types G5 to M5
(after elimination of one particularly uncertain luminosity class 
and one known metal poor star with obviously weaker spectral features). 
Their shortest wavelength is 0.97\,$\mu$m, except for 5 spectra
with an optical extension.

\begin{figure}
\includegraphics[clip=,width=0.49\textwidth]{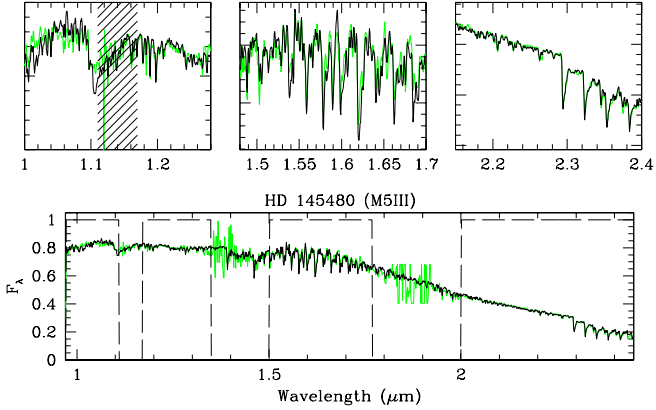}
\caption[]{Fit to a late type giant star spectrum (HD\,145480, M5III). 
The data are shown as dotted lines, the model as a solid line. The
window function for the $\chi^2$ fit is also shown.
Model parameters are 3400\,K, log($g$)=0, A$_V$=0.4, $\chi^2=1.7$.
Such a fit quality is typical for all the giants without available
optical spectra.}
\label{goodfit_M5III.fig}
\end{figure}

\begin{figure}
\includegraphics[clip=,width=0.48\textwidth]{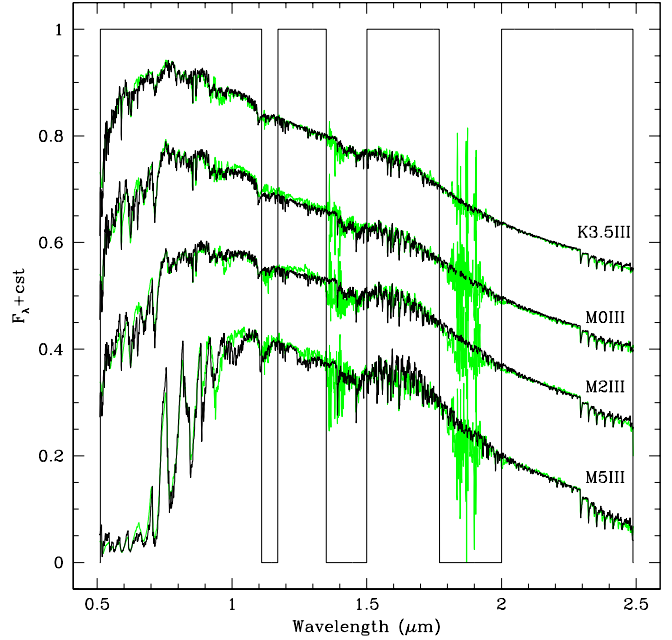}
\caption[]{Best fits to four giant star spectra 
that extend to optical wavelengths.
Data are shown as dotted lines, best fit models as solid lines. 
The window function used to reject the noisiest spectral regions is also shown.
From top to bottom: K3.5III star BS\,4432 with model
[\Teff, log($g$), A$_V$, $\chi^2$] = [4100\,K, 2, 0.55\,mag, 2.3];
M0III star BS\,4371 with [3900\,K, 1, 0.2\,mag, 4.5];
M2III star BS\,5301 with [3800\,K, 1, 0.55\,mag, 5.2];
M5III star BS\,4267 with [3300\,K, 1, -0.3\,mag, 13.9].
%
}
\label{fits_4VKgiants.fig}
\end{figure}

Good fits are obtained for essentially all the near-IR spectra
with the solar metallicity models. An example is given 
in Fig.\,\ref{goodfit_M5III.fig}. Among the satisfactory fits, there
is a tendency for $\chi^2$ to increase systematically  
from about 1 for types $\leq$\,K4 
($\chi^2$=0.6-1.4, depending on the actual S/N of individual spectra) 
to about 2 (1.5-2.5) for the M stars. This trend is due to a wealth of
weak lines and deeper molecular bandheads at low temperatures, which among
others induces a higher sensitivity of the value of the $\chi^2$
to residual wavelength calibration errors or model line lists. 

The $\chi^2$ values for combined optical+near-IR
spectra take values of 3 to 7 for satisfactory fits (considering the sensitivity
of the $\chi^2$ to the smoothing parameter at optical wavelengths and
to flux calibration errors over such an extended wavelength range).
Examples are shown in Fig.\,\ref{fits_4VKgiants.fig}.
The best fit to the spectrum of the coolest giant, 
the M5III star BS\,4267 (=\,HD\,94705) requires 
marginally negative extinction. Considering the flux calibration uncertainties
and the choice of one particular extinction law, such a result is
not alarming. Cooler models by only 100\,K or models with higher
log($g$) by 0.5 would result in a positive value of the estimated A$_V$.
The most obvious shortcoming of the models for giants this cool is
an overprediction of the TiO absorption bands near 1 and 1.25\,$\mu$m.

The results of the fitting procedure can be summarized as follows: 

Temperatures range from 5300\,K for type G5, to 3300\,K for type M5.
As expected from stellar evolution tracks, the highest 
available gravity (log($g$)=2) is selected for giants
earlier than K7 (with one exception), 
then progressively more spectra are assigned 
log($g$)=1 and later 0. A$_V$ values are spread between 0 and 1
(with 4 cases of marginally negative values). No correlation is found 
between A$_V$ and \Teff .


\medskip

Adopting the models with RSG-specific abundances rather than solar ones
leads to poorer fits in all but one case. The values of the reduced $\chi^2$
increase by 0.5 units on average. While the distribution of estimated 
effective temperatures for the sample is relatively flat with the assumption
of solar abundances, it becomes bimodal with RSG-specific abundances\,:
temperatures between 4500 and 5000\,K are systematically avoided, because
the CN bands of these models are too strong at the surface gravities of 
giant stars (cf. Fig.\,\ref{Teff_CN.fig}).
This result was expected from Figs.\,\ref{data_indices.fig} and
\ref{data_indices_colours.fig}. 
It is satisfactory as our set of RSG-specific abundances 
is not designed to match abundances in giant stars.



\subsection{Bright giants}

\begin{figure}
\includegraphics[clip=,width=0.48\textwidth]{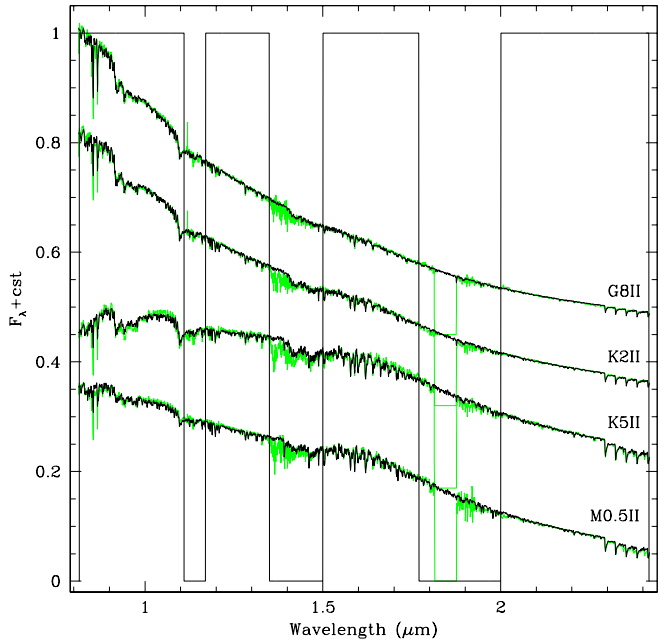}
\caption[]{Best fits to four bright giant star spectra (class II) 
that extend down to 0.81\,$\mu$m. Figure set-up is as in 
Fig.\,\ref{fits_4VKgiants.fig}.
From top to bottom: G8II star HD\,150416 with
[\Teff, log($g$), A$_V$, $\chi^2$] = [5000\,K, 1, 0.1\,mag, 0.65];
K2II star BD-29\,2374 with [4500\,K, 2, -0.15\,mag, 0.65];
K5II star HD\,168815 with [4100\,K, 1, 1.3\,mag, 1.8];
M0.5IIb star HD\,132933 with [4000\,K, 2, 0.4\,mag, 1.4].
Among the above, we classify the K5II fit as satisfactory, the others
as very good.
}
\label{fits_4TKbrights.fig}
\end{figure}

\begin{figure}
\includegraphics[clip=,width=0.48\textwidth]{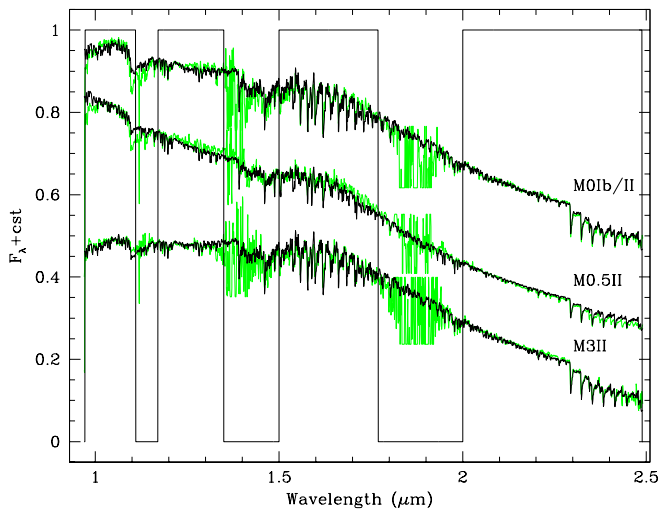}
\caption[]{Best fits to two bright giant star spectra for which the 
CN band at 1.1\,$\mu$m is particularly poorly reproduced, and of the
coolest class II star of the sample.
Figure set-up is as in Fig.\,\ref{fits_4VKgiants.fig}.
From top to bottom: M0Ib/II star HD\,145384 with
[\Teff, log($g$), A$_V$, $\chi^2$] = [3400\,K, -0.5, -0.15\,mag, 3.2];
M0.5II star HD\,142676 with [3900\,K, 2, 0.0\,mag, 4.0];
M3II star HD\,153961 with
[3500\,K, 0, 1.0\,mag, 3.3].
}
\label{fits_3IKbrights.fig}
\end{figure}

The sample contains 29 bright giants of class Ib/II or II. 
Spectral types range from G3 to M3. None of the
corresponding spectra extend through optical wavelengths, but 11 extend
down to 8100\,\AA. Their properties in terms of colours and molecular
indices are spread between those of giants and supergiants.
On average, they display slightly stronger bands of CO
and significantly stronger bands of CN than giants of class III, at
a given (dereddened) colour.

The solar metallicity model fits to all the spectra
are satisfactory, two thirds of them being very good
(Fig.\,\ref{fits_4TKbrights.fig}).
The models clearly contain all the molecular bands required. 
Marginally negative values of A$_V$ are obtained in four cases, which
is again not unexpected considering flux calibration uncertainties.
The most common shortcomings found when the fits are not perfect
are the following: \\
--- There is a tendency for the models to show stronger CO bands at 
2.29\,$\mu$m and weaker CN bands at 0.93 and 1.1\,$\mu$m than observed. 
This problem is only detectable clearly 
when the data extend down to 0.81\,$\mu$m.\\
--- For stars with spectral types around K5II whose spectra extend down
to 0.81\,$\mu$m, the models struggle to reproduce the energy distribution
around 1\,$\mu$m, where it peaks between deep CN bands. This difficulty
is certainly related to the strength of the CN bands at those temperatures
(see Fig.\,\ref{Teff_CN.fig}). \\
--- In two cases (spectral type M0Ib/II and M0.5II), the model CN bands 
are too weak while CO is reproduced well (Fig.\,\ref{fits_3IKbrights.fig}).

The \Teff\ and log($g$) scales obtained for the bright giants have a 
scatter similar to those found for giants and supergiants and are located
between the two, as expected. Bright giants with spectral types earlier
than K4 (included) are assigned log($g$)=2 (with one exception\,:
HD\,170457, G8Ib/II), and values of log($g$) for types K5-M3 are 
scattered between 1 and -0.5. No correlation is found between A$_V$ and 
\Teff .


\medskip

When moving from solar abundances to RSG-specific abundances, 
the $\chi^2$ test indicates that the fits are degraded in a majority
of cases (typically by 0.5 $\chi^2$ units, as for the sample of 
class III stars). However, a significantly improved $\chi^2$ is obtained 
for 4 stars, and the $\chi^2$ changes are insignificant in 7 cases.  \\
The improvements correspond to four stars of type K2 to M0 
(out of the 11 bright giants available in this range), 
with estimated \Teff\ of 4300 to 3400\,K.
%
%
Eye inspection of the corresponding four spectra shows that the decrease
in $\chi^2$ corresponds to a better fit to the CN bands, which were not
deep enough (by small amounts) in the solar metallicity models.
In some cases, the improved $\chi^2$ was associated with a decrease in \Teff\
by 100\,K or an increase of log($g$) by one bin size, but statistics are
too small to define significant trends.\\
Degraded fits are frequently associated with excessive strengths of the 
model CN bands when the RSG-specific abundances are used. 
The \Teff -distribution obtained with the RSG-specific abundances still
shows a zone of avoidance between 4500 and 5000\,K, but the effect is
not as obvious as in the case of class III stars. Although small number
statistics affect this result, we note that all class II spectra with
estimated \Teff\ in that range have poorer fits with the RSG-specific abundances
than with solar ones.  As expected, models with the adopted RSG-specific
abundances do not apply to the majority of class II stars but they
do allow us to identify candidate objects that may have suffered 
more than standard dredge-up.

\subsection{Supergiant stars}
\label{supergiants.sec}

\begin{figure}
\includegraphics[clip=,width=0.48\textwidth]{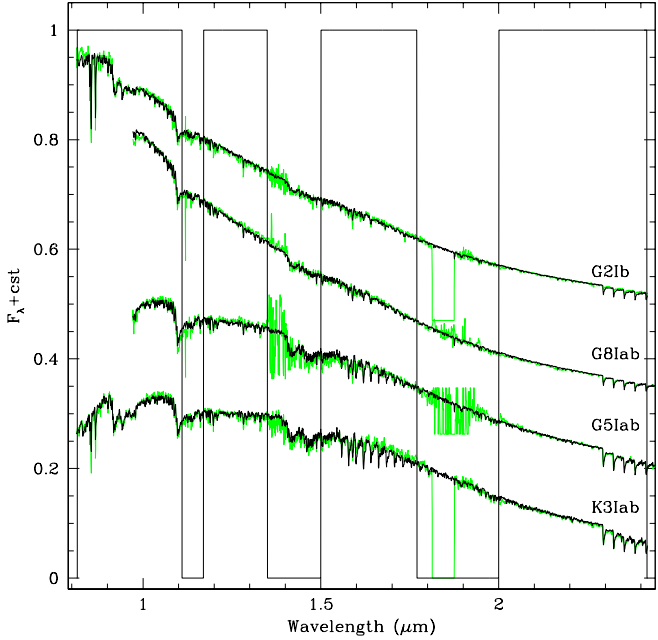}
\caption[]{Good and reasonable best-fits 
to warm red supergiant spectra (class I).
Figure set-up is as in
Fig.\,\ref{fits_4VKgiants.fig}.
Abundances are solar.
From top to bottom: G2Ib star HD\,182296 with
[\Teff, log($g$), A$_V$, $\chi^2$] = [5000\,K, 0, 0.95\,mag, 1.3];
G8Iab star HD\,206834 with [4900\,K, 1, 0.3\,mag, 0.85];
G5Iab star HD\,170234 with [4500\,K, 0, 2.15\,mag, 1.8];
K3Iab star HD\,187238 with [4100\,K, 0, 1.65\,mag, 2.2].
We have not counted the K3Iab case as a good fit, because the model
CO bands around 1.7\,$\mu$m are too strong. Note that the spectral type
of the second and third stars are likely to be incorrect.
}
\label{fits_goodsuper.fig}
\end{figure}

\begin{figure}
\includegraphics[clip=,width=0.48\textwidth]{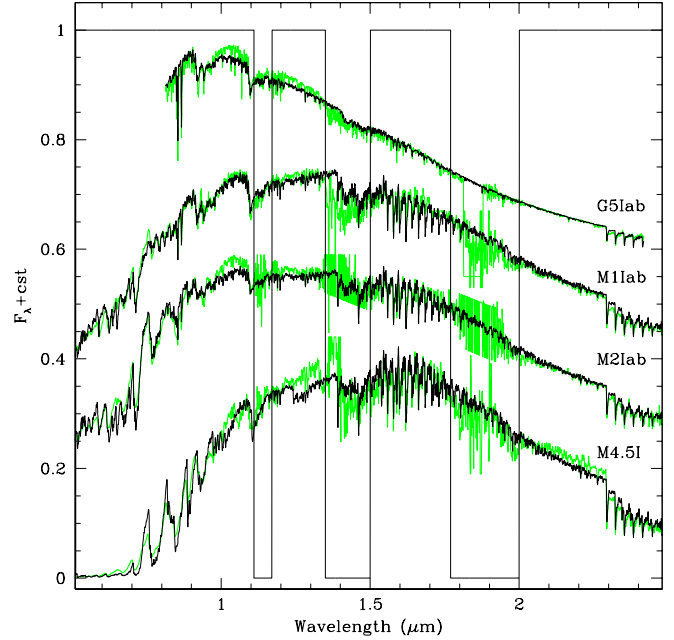}
\caption[]{A selection of marginally acceptable best-fits to red supergiant
spectra.  Figure set-up is as in
Fig.\,\ref{fits_goodsuper.fig}.
From top to bottom: G5Iab star HD\,165782 with
[\Teff, log($g$), A$_V$, $\chi^2$] = [4900\,K, $-1$, 2.4\,mag, 2.8];
M1Iab star HD\,98817 with [3700\,K, $-1$, 1.8\,mag, 7.8];
M2Iab star BS\,3364 (=\,HD\,72268) with [3500\,K, $-0.5$, 1.0\,mag, 12.4];
M4.5I star V774\,Sgr with [3200\,K, $-0.5$, 1.7\,mag, 22.7].
}
\label{fits_medsuper.fig}
\end{figure}

\begin{figure}
\includegraphics[clip=,width=0.48\textwidth]{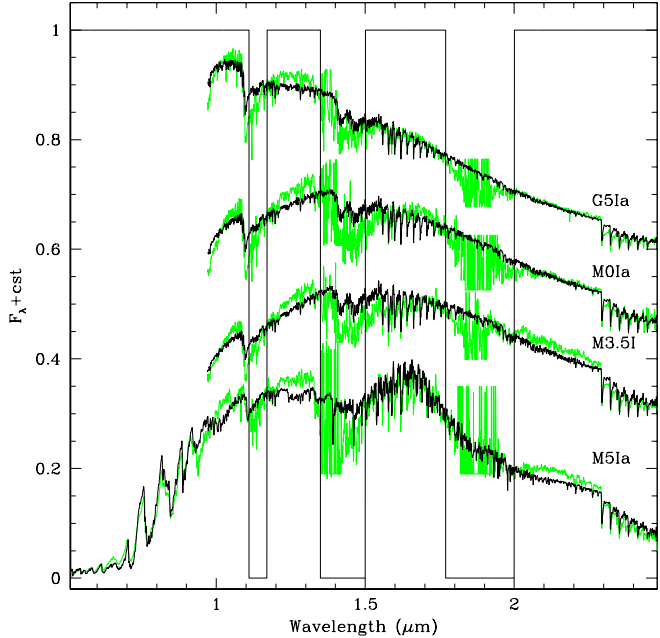}
\caption[]{A selection of poor best-fits to red supergiant
spectra.  Figure set-up is as in
Fig.\,\ref{fits_goodsuper.fig}.
From top to bottom: G5Ia star HD\,155603 
(classified K0\,0-Ia by Keenan and McNeil, 1989)
with [\Teff, log($g$), A$_V$, $\chi^2$] = [4300\,K, $-0.5$, 1.8\,mag, 8.6];
M0Ia star Trumpler\,1-27 with [4200\,K, $-1$, 4.4\,mag, 13.4];
M3.5I star IRC\,$-$20427 with [4000\,K, $-1$, 5.35\,mag, 15.2];
M4-5Iab star CL\,Car with [3300\,K, $2$, 1.65\,mag, 21.6].
The fits being of poor quality, the derived parameters are not 
reliable and are given here only for the sake of completeness.
}
\label{fits_poorsuper.fig}
\end{figure}

The data sample contains 37 spectra of stars of class I, Ia, Iab or Ib
(after removal of one particularly odd case that is probably misclassified
and of one spectrum with poor correction for telluric absorption). Spectral
types range from G2 to M5. The stars with the latest spectral types 
($\geq$\,M2) are all known or suspected variables (as are the vast majority
of late type supergiants in nature). 9 spectra, all with spectral type M,
extend through the optical range; note that the optical and near-IR spectra 
of individual objects
were taken within less than three weeks of each other. 8 spectra extend
down to 0.81\,$\mu$m.

Good fits to the red supergiant spectra with solar metallicity models
are obtained for 13 of the 37 spectra, all of which are of 
spectral type G2-G8. 16 of the remaining spectra find a model representation
that is still reasonable (though often significantly poorer than the fits
we called satisfactory within class II above).
These are spread over the whole range of 
spectral types and include some of the data that extend through optical
wavelengths. In general, stars of luminosity class Ib are easier 
to fit than those of class I, Ia or Iab, and all class Ib stars of 
our sample can be matched well or reasonably well.
Finally, we classify 7 of the red supergiant fits as poor. Five of these
correspond to variable stars with spectral types later than M3.5 (class
I, Ia or Iab), the two others are of spectral type M0Ia and G5Ia.

Figures \ref{fits_goodsuper.fig}, \ref{fits_medsuper.fig} 
and \ref{fits_poorsuper.fig} 
illustrate some of the good, intermediate and poor model
fits. 

The main shortcomings of the models when the fits are 
of {\em intermediate} quality are the following\,: \\
--- A relatively common feature is a shortage of flux in the models 
around 1\,$\mu$m, as seen in two spectra of Fig.\,\ref{fits_medsuper.fig}.
This problem was already mentioned for a few bright giants of class II,
as a property that is associated with strong CN bands and can be identified
only when the observed spectra extend far enough to short wavelengths. \\
--- Even when the 1st overtone CO bands after 2.29\,$\mu$m are reproduced 
reasonably
well, it happens that the relative strengths of the 2nd overtone CO bands in the
H window are incorrect, the transitions at longer H window wavelengths 
(1.65-1.75\,$\mu$m) being too strong in the models compared to the data 
(last spectrum of 
Fig.\,\ref{fits_goodsuper.fig} and 2nd and 3rd spectrum of 
Fig.\,\ref{fits_medsuper.fig}). \\
--- In the coolest models, bands of TiO appear (in particular near 1.25\,$\mu$m)
that are not seen in the data. 
\medskip

{\em Poor} fits are obtained for the coolest stars
(e.g. bottom spectrum of Fig.\,\ref{fits_poorsuper.fig}) 
and for stars with extreme CN bands 
(e.g. top three spectra of Fig.\,\ref{fits_poorsuper.fig}). 
We recall that the coolest stars are also variable 
and that discrepancies
are to be expected in a comparison with static models. 
When the CN bands are strong, the derived temperatures are a compromise
between the necessity to reproduce the energy distributions and the CO
bands at 2.29\,$\mu$m
(which pulls towards low temperatures), and the need to maximize CN depths
(which pulls towards 4100\,K, cf. Fig.\,\ref{Teff_CN.fig}). When optical
spectra are taken into account, the relative weight of the CN 
bands is reduced compared to CO, optical features and the energy
distribution. On the contrary, when only wavelengths between 0.97
and 2.4\,$\mu$m are available the r\^ole of the CN bands is large.
This explains why in Fig.\,\ref{fits_poorsuper.fig} such a large
difference in \Teff\ is obtained between the M3.5Ia star (no optical
spectrum, best fit \Teff =\,4000\,K) 
and the M4-5I star (optical spectrum available, best fit \Teff =\,3300\,K). 
The temperatures of the M0Ia and M3.5I stars 
of that figure are most probably overestimated. For a similar reason,
the temperature of the G5Ia star at the top of the figure may be 
underestimated (compare with the G5Iab star in Fig.\,\ref{fits_medsuper.fig}).

A typical problem with the best fit models for the spectra with 
very strong CN is the relative strength of the various CO bands.
The G5Ia star HD\,155603 (Fig.\,\ref{fits_poorsuper.fig}) provides the
most extreme example. It has the strongest 2.29\,$\mu$m CO band of our 
whole supergiant sample and among the strongest CN bands as well,
but in the H window CO is almost absent. 
None of the current models with
$v_{\rm mic}$\,=\,2\,km\,s$^{-1}$ 
reproduces this combination. Models with larger micro-turbulent 
velocities improve the representation of these extreme spectra.

Water bands are another cause of disagreement between models and data.
The near-IR bands of H$_2$O and CN  overlap in wavelength
to such a degree that confusion
can occur (and has occurred in the past, cf. Wing \& Spinrad 1970). The
shapes of the 1.1\,$\mu$m bands of H$_2$O and CN are subtly different
(cf. Fig. 5 of Lan\c{c}on \& Wood 2000). The bands observed in red 
supergiants correspond closely to CN, although contamination with 
H$_2$O is possible. The H$_2$O band around 1.9\,$\mu$m,
which is very deep and broad in Miras, is inconspicuous in red supergiants.
It may be present at a low level in the coolest ones observed, 
such as CL\,Car (Fig.\,\ref{fits_poorsuper.fig}), which are semi-regular 
long period variables. 
The clearest H$_2$O signature 
in the observed red supergiant spectra is a sharp bandhead at 1.32\,$\mu$m,
although the detection of this feature requires good corrections for
telluric absorption bands. Based on this signature, the 7
coolest supergiants of our sample contain H$_2$O (all these are variable). 
The models however either do not 
show this bandhead (low $g$) or, when they do (high $g$), 
also display a 1.9\,$\mu$m band 
that is much wider and deeper than observed.

Finally, the semi-regular variables V774\,Sgr, EV Car and CL Car
(Figs.\,\ref{fits_medsuper.fig} and \ref{fits_poorsuper.fig})  
have a clear VO absorption band at 1.05\,$\mu$m and small or inexistent
absorption bands at 1\,$\mu$m 
and 1.25\,$\mu$m,
two properties that are not matched by the models.


\medskip

\begin{figure}
\includegraphics[clip=,width=0.48\textwidth]{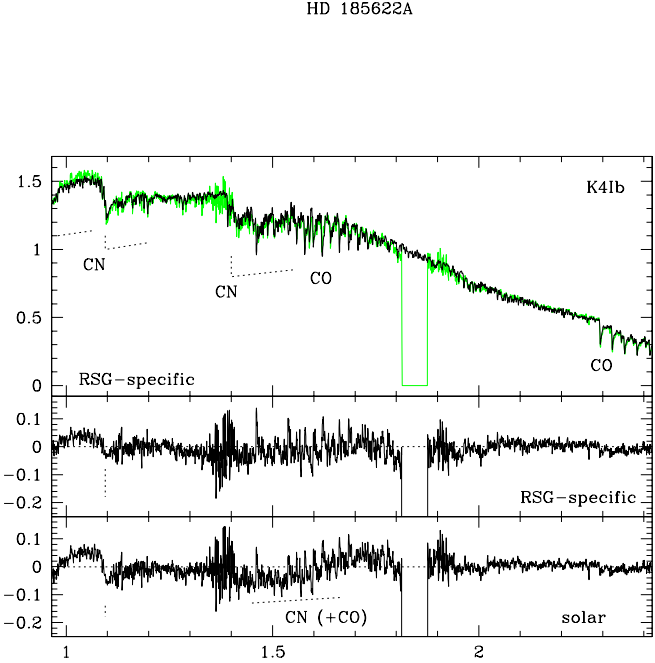}
\caption[]{{\em Top:}\ K4Ib star HD 185622a and best fit model with 
RSG-specific abundances. {\em Middle:}\ Residuals of the fit shown above
(data$-$model). {\em Bottom:}\ Residuals of the best fit with solar
metallicity models ($\chi^2$ 1.22 times larger than in the RSG-specific case). 
Note the CN residuals below 1\,$\mu$m, around 1.1\,$\mu$m and in the 
slope between 1.45 and 1.75\,$\mu$m. These are typical and systematic
shortcomings of the solar metallicity models in the cases where 
RSG-specific models provide a better fit.}
\label{residuals.fig}
\end{figure}

When moving from the models with solar abundances to the RSG-specific 
abundances, the $\chi^2$ test indicates that about a third of the fits
are improved, another third are degraded, and the quality of the final 
third of the fits is essentially unchanged. 

The deteriorations, when present, are not severe. In most cases, it seems that 
abundance values intermediate between the adopted solar and RSG-specific
sets would provide optimal fits, which is not surprising considering
that evolutionary tracks for red supergiants cover a range of abundances.
Eye inspection shows that quite a few stars with equally 
good fits with both model sets also fall in this category.

The improvements obtained with RSG-specific abundances for a fraction
of the red supergiants are significant,
although they clearly do not resolve all the difficulties. 
They are associated with a better representation 
of the observed CN bands and sometimes also with a better match to the CO bands
around 1.6\,$\mu$m (see also Sect.\,\ref{weights.sec}). 
One may distinguish two subcategories
of improvements. On one hand, some of the stars that already 
had reasonable model counterparts with solar abundances 
have better, often good adjustments with RSG-specific abundances. 
These are mainly stars of type G and K. An example is given in
Fig.\,\ref{residuals.fig}.
On the other hand, the improvements refer to stars that had poor fits
with solar abundances, and for which the RSG-specific abundances 
lead to somewhat better but still unacceptable fits. These are the
same 7 stars as mentioned earlier. The models cannot simultaneously reproduce
their CO bands (1.6 and 2.29\,$\mu$m), their CN bands and 
their energy distribution. More extended model grids are 
needed to characterize these objects. Problems related to H$_2$O, TiO and VO,
when present, remain essentially unchanged. 

The explored changes in surface abundances induce changes
in the best-fit parameters for the sample of observed stars with
maximal amplitudes of $\pm$200\,K. For the sample as a whole,
there is no strong correlation between the change in \Teff\ 
and the actual value of \Teff,
which is not surprising considering that many fits are imperfect
and that the behaviour expected from theory is complex
(see Sect.\,\ref{Teffscales_models.sec}). 

The \Teff\ distribution of the red supergiant sample
obtained under the assumption of RSG-specific abundances
shows no anomaly. Scrutiny of the 2D distribution of estimated
parameters in the log($g$)--\Teff\ plane suggests that a narrow zone  
extending diagonally from [log($g$)=0,\Teff =4000\,K] to
[log($g$)=1,\Teff =5000\,K] (with no extension to lower gravities)
might nevertheless be underpopulated. 
The statistical significance of this gap is low because of small 
sample numbers. Its presence would favour a general
picture in which RSG-specific abundances are only relevant to red 
supergiants with large initial masses or to late stages
of red supergiant evolution.

\subsection{Effects of the weighting of various parts of the spectra}
\label{weights.sec}

\begin{figure}
\includegraphics[clip=,angle=270,width=0.49\textwidth]{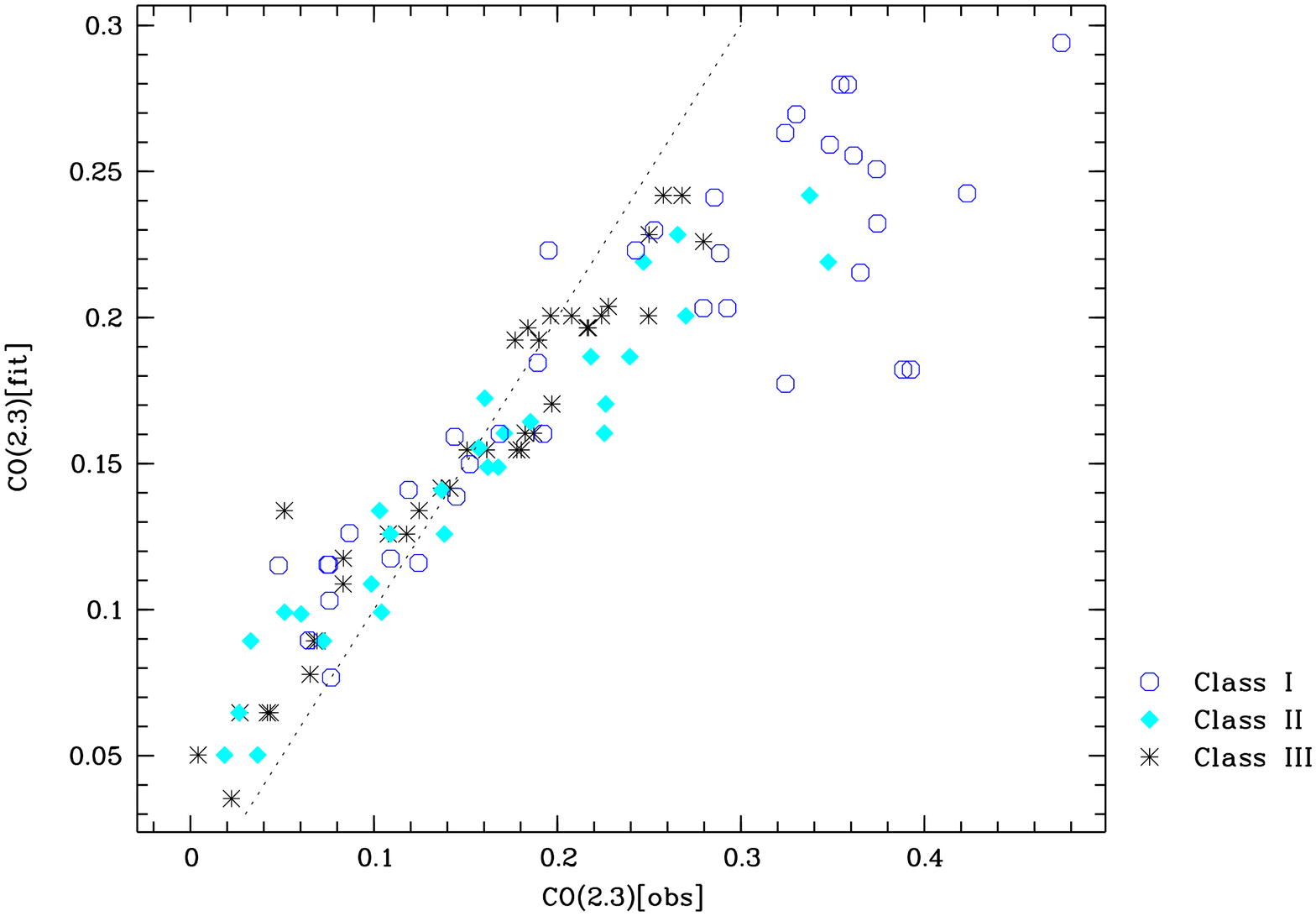}
\includegraphics[clip=,angle=270,width=0.49\textwidth]{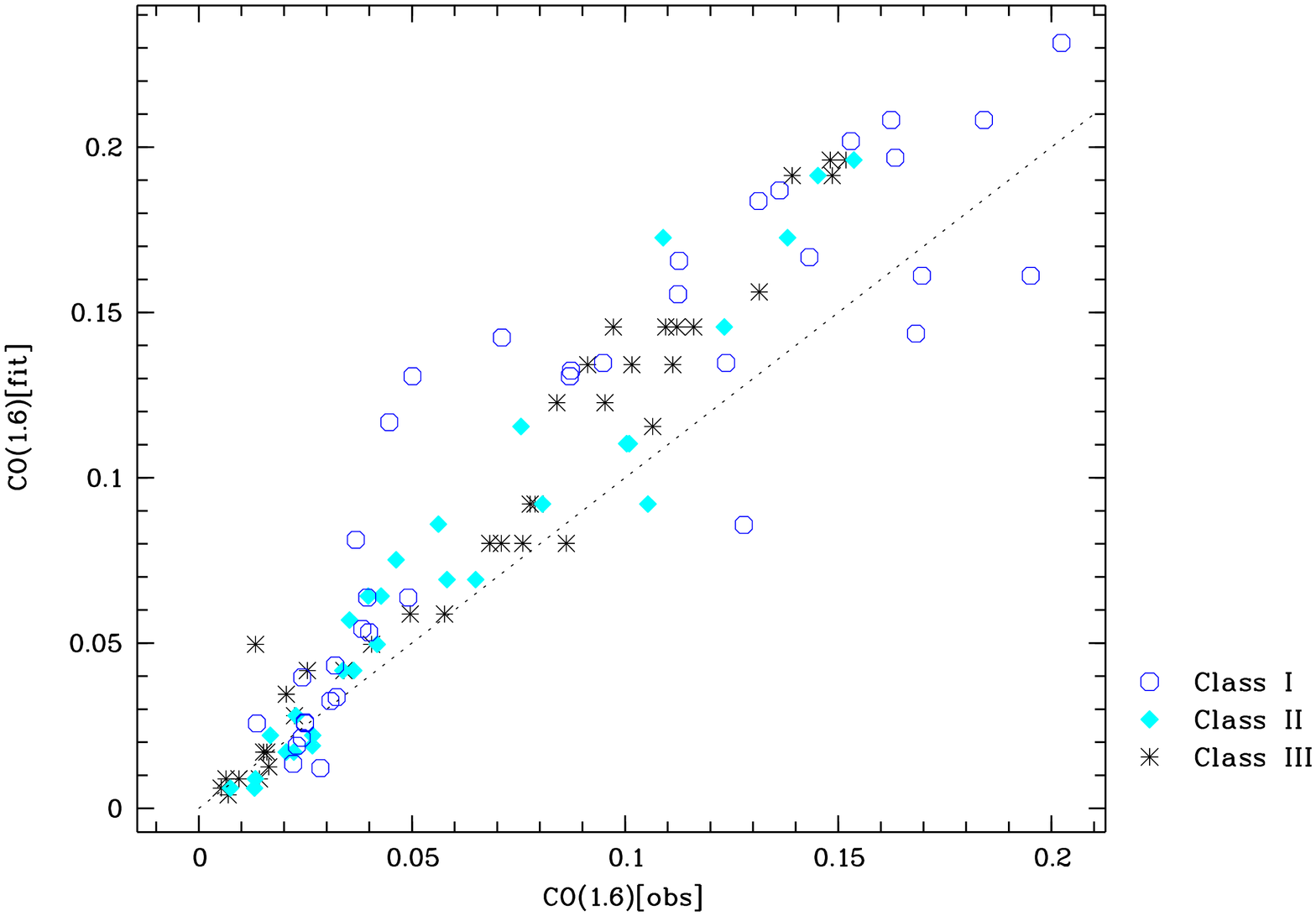}
\includegraphics[clip=,angle=270,width=0.49\textwidth]{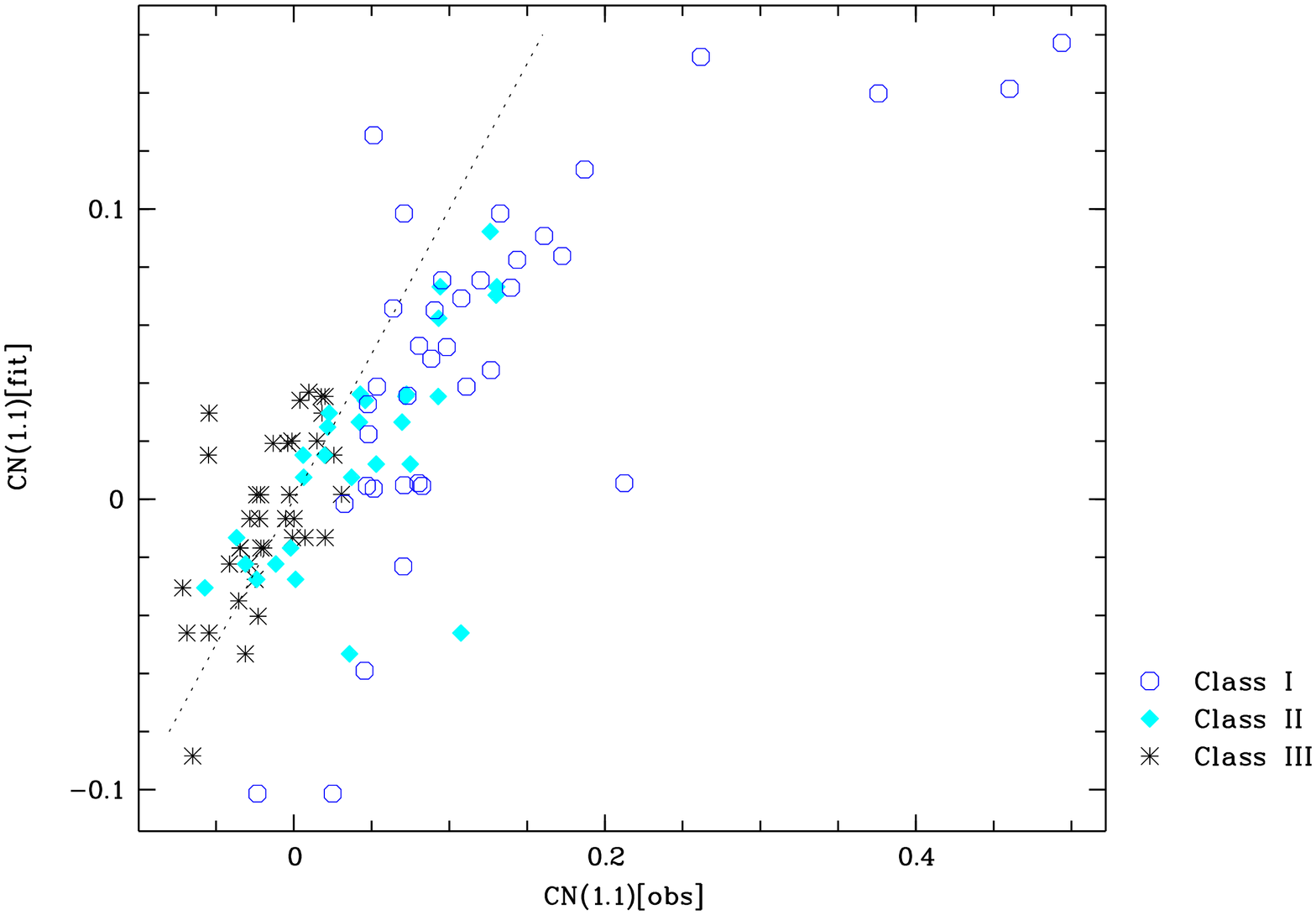}
\caption[]{{\em Top:}\ Strength of the 2.3\,$\mu$m CO band of the 
best fitting solar metallicity
models versus strength of this band in the dereddened observed
spectra (cf. Tab.\ref{indexdef.tab}). 
The dotted line highlights the one-to-one relation.
{\em Middle:}\ Same figure for the adopted measure of the 1.6\,$\mu$m CO band.
{\em Bottom:}\ Same figure for the adopted measure of the 1.1\,$\mu$m CN band.
}
\label{fitquality.fig}
\end{figure}

Because there are {\em systematic} differences between the best fit models
and the observed spectra, the best fit model parameters depend on 
the weights given to the various spectral features in the fitting 
procedure. Our standard method weights the data based on a
reasonable simplified model for the high frequency noise in the data.
This adopted weight is inversely proportional to the square root of 
the signal, i.e. spectral regions with large fluxes contribute more to
the $\chi^2$ than regions with small fluxes. Since the spectra of cool
stars peak around 1\,$\mu$m (in the flux density units adopted in this paper),
molecular bands near this wavelength are important in the fit. 
In practice, this weighting makes CN bands relatively important and 
CO bands (around 1.6\,$\mu$m and 2.3\,$\mu$m) comparatively unimportant.

If the noise was indeed gaussian, uncorrelated between pixels, and 
exactly of the amplitude assumed, then our procedure would select the
models with the largest likelihoods. This is not the case (flux calibration
errors, wavelength-dependent gains and contributions of the read-out noise, 
etc.), and therefore our weighting is in some ways non-optimal. We may
choose various alternative methods (see Decin et al. 2004 for an 
interesting discussion of comparable issues). First, we may decide to fit  
measurements of the depths of one or several features rather than spectral
segments. Unfortunately, the selection of one or the other feature remains
somewhat arbitrary. Second, we may decide to focus on either the optical or the
near-IR spectral range. This circumvents the difficulty
of reproducing the global energy distribution (possible uncertainties
in the relative flux calibration of the optical and near-IR data, uncertainties
in the adopted extinction law, etc.). Third, we may keep the whole spectrum
but explore the effects of alternative weightings. We briefly
summarize below the main trends found while investigating these three options.
\medskip

\begin{figure}
\includegraphics[clip=,angle=270,width=0.49\textwidth]{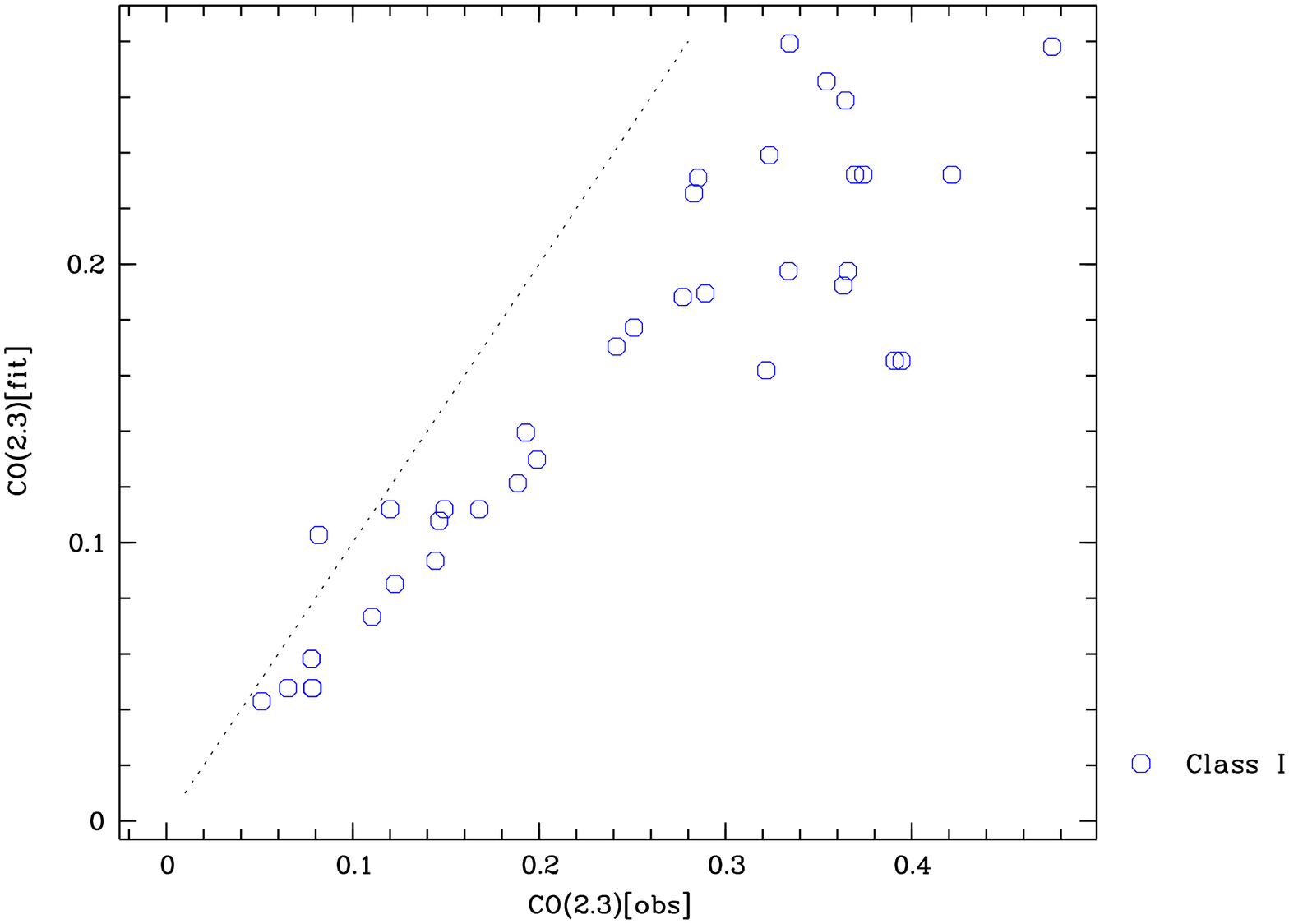}
\includegraphics[clip=,angle=270,width=0.49\textwidth]{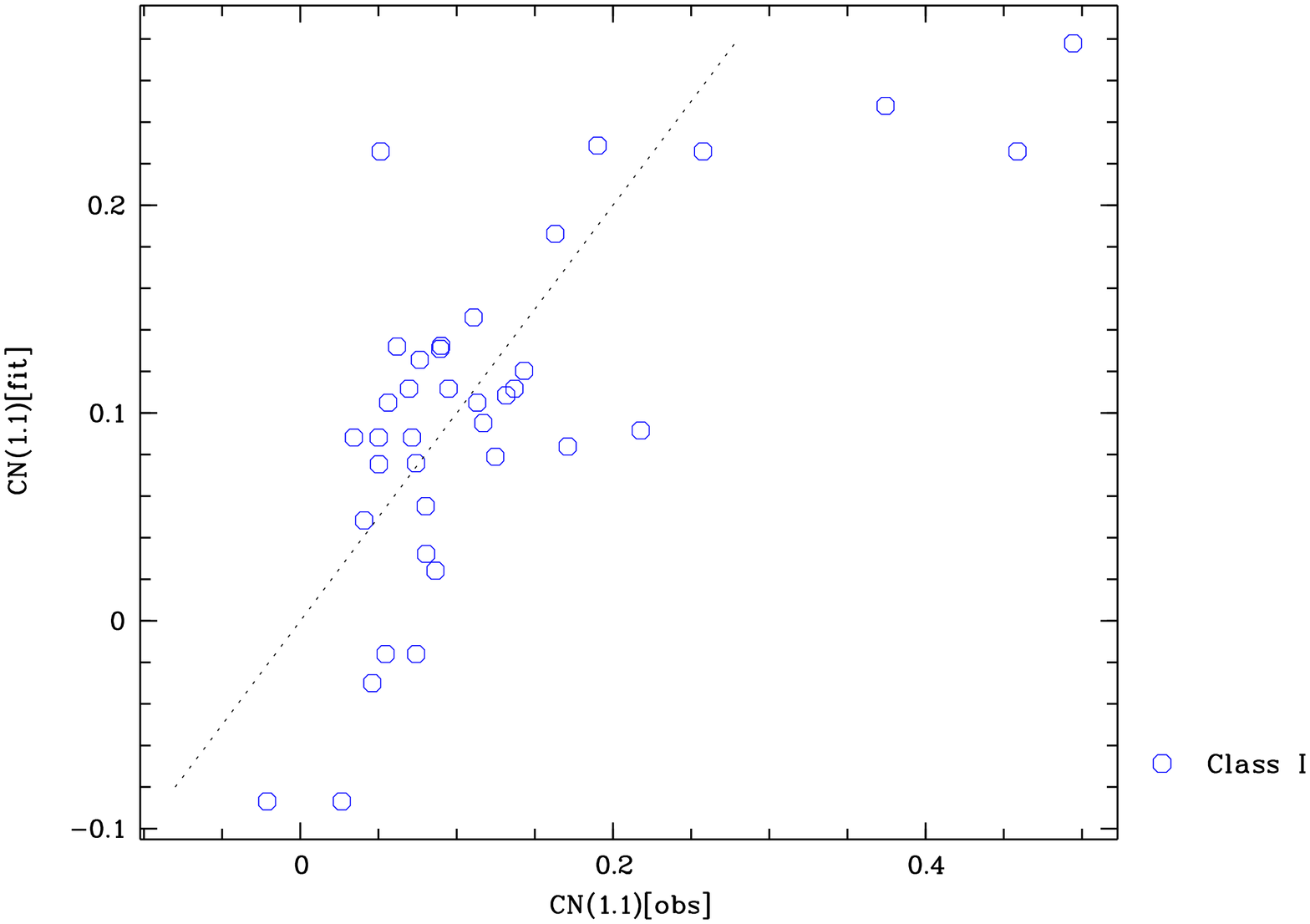}
\caption[]{Same as the top and bottom panels of Fig.\,\ref{fitquality.fig} 
but for supergiants only, and using models with RSG-specific abundances.}
\label{fitquality_set6.fig}
\end{figure}

Figures \ref{fitquality.fig} and \ref{fitquality_set6.fig}
show how three important near-IR molecular features in the models compare 
with their observed counterparts, when our standard
weighting procedure is applied to select the best fit.
Note that the corresponding figures for H-K, J-H, J-K or 104-220 (not shown) 
are very well behaved, which only states that the adopted extinction law
is capable of dealing with the actual extinction (and with 
flux calibration errors) rather well.

As expected, systematic drifts away from the perfect match are smaller
for the 1.6\,$\mu$m CO features than for the 2.3\,$\mu$m CO band, the
latter being located in a region of low flux (small contribution to
the $\chi^2$). The best-fit models have systematically deeper 2.3\,$\mu$m CO
bands than the data for warm stars (types G and K), but systematically
too shallow 2.3\,$\mu$m CO bands for the coolest stars (type M).
By changing the weights in the fitting procedure (e.g. by assuming 
a constant signal-to-noise ratio), the 2.3\,$\mu$m bands can be reproduced
better, but at the cost of a loss of the fit quality at shorter 
near-IR wavelengths.

The CN bands are reproduced well for giant stars. But they are too 
shallow in the best fit models for some of the bright giants 
and for the supergiants.
Here, changing the fitting weights has a small effect compared to
more fundamental model parameters such as abundances, gravities or 
micro-turbulence.  RSG-specific abundances 
move the bulk of the red supergiants into a satisfactory location
(Fig.\,\ref{fitquality_set6.fig}). With RSG-specific abundances,
the fits to CO bands around 1.6\,$\mu$m are not fundamentally improved or
degraded on average, while the first overtone CO bands (2.3\,$\mu$m)
of the best fits become shallower, i.e. too shallow. 
By assigning CO more weight in the fits, it is
possible to reduce this discrepancy while still observing the global
improvement for CN. But with the current grid of models, no
fully satisfactory solution can be found for any weighting scheme.    
\medskip

The weights given to various spectral ranges impact on the estimated
stellar parameters. Examples have already been given in 
Sect.\,\ref{supergiants.sec}, and further discussions can be found below. 



\section{Discussion}
\label{discussion.sec}

Providing estimates of fundamental stellar parameters is a major
application of theoretical spectra. Our discussion focuses 
on the determination of \Teff\ from near-IR spectra using the 
new \Phoenix\ models.

\subsection{Stellar mass}
\label{discussion.mass.sec}


We have mentioned in Sect.\,\ref{trends.sec} that the effects of mass
on colours and molecular indices, at a given \Teff\ and log($g$), are small.
The comparison between the
best fit parameters obtained assuming 1\,M$_{\odot}$ and 15\,M$_{\odot}$
nevertheless reveals a trend worth highlighting : {\em for stars  with
high surface gravities (log($g$)=2), the temperatures obtained
with 1\,M$_{\odot}$ models are systematically lower by $\sim$100\,K
than those obtained with 15\,M$_{\odot}$ models.} This is particularly
relevant to giants of class III, but also to the warmer of the
bright giants of class II.
Unfortunately, we found no systematic differences between the
$\chi^2$ values obtained with one or  the other assumption on mass.
Thus, {\em it is not currently possible to determine mass using spectral fits}
such as those performed in this paper.
Mass has to be fixed a priori by other means.

For luminous giants and supergiants, i.e. stars with low gravities,
we found no systematic effect of mass on best-fit \Teff\ or log($g$).
The differences in \Teff\ between the two assumptions are scattered
around 0 with typical values of $\pm$100\,K (more in cases where
even the best fits are not satisfactory). We note a correlation between
the difference in \Teff\ and the difference in log($g$)\,: when a change
in the assumed mass leads to a rise in the best-fit \Teff, it generally
also produces a rise of the best-fit value of log($g$).

\subsection{\Teff -systematics related to surface abundances: model 
predictions}
\label{Teffscales_models.sec}

\begin{figure}
\includegraphics[clip=,width=0.49\textwidth]{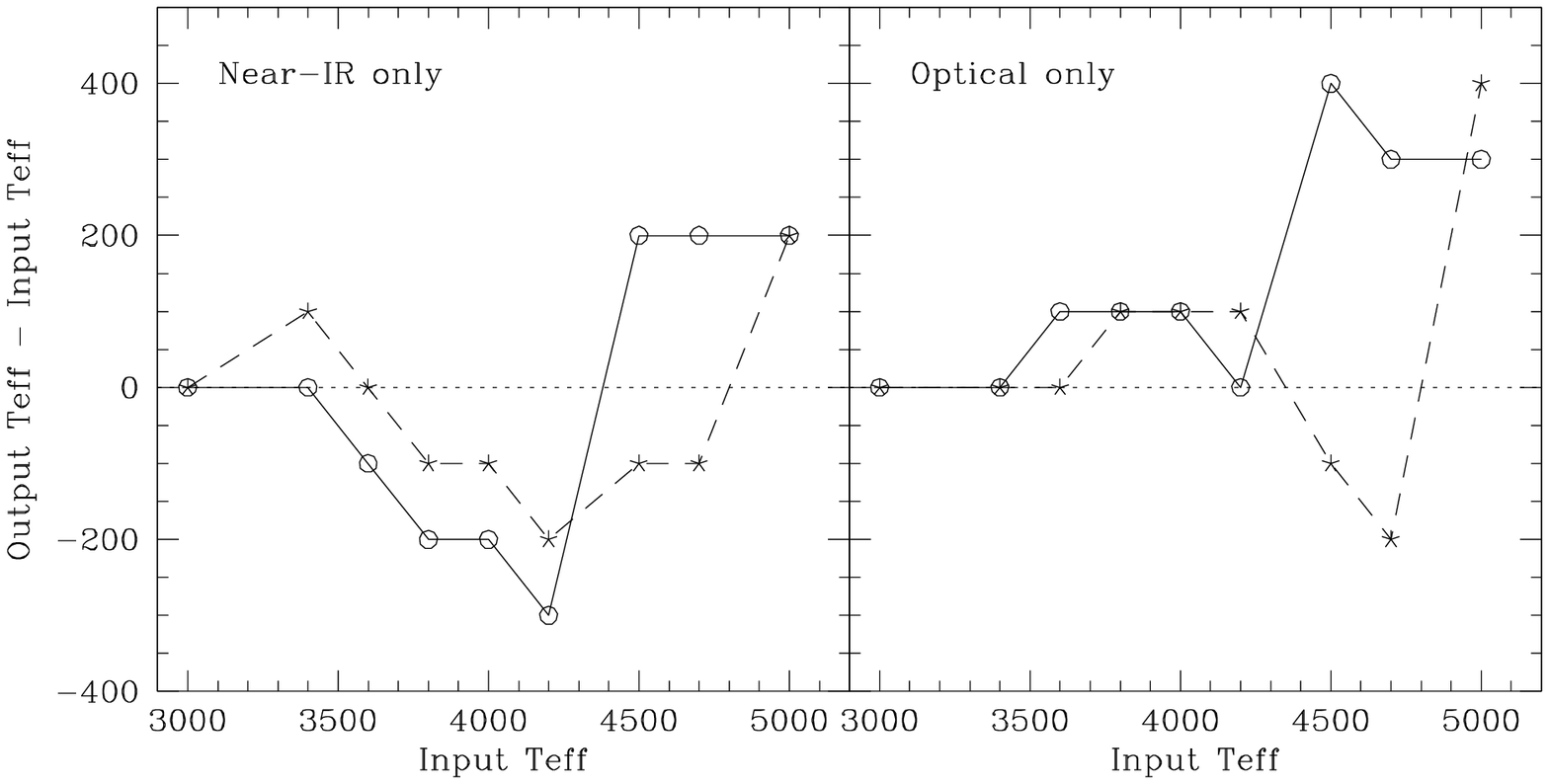}
\includegraphics[clip=,width=0.49\textwidth]{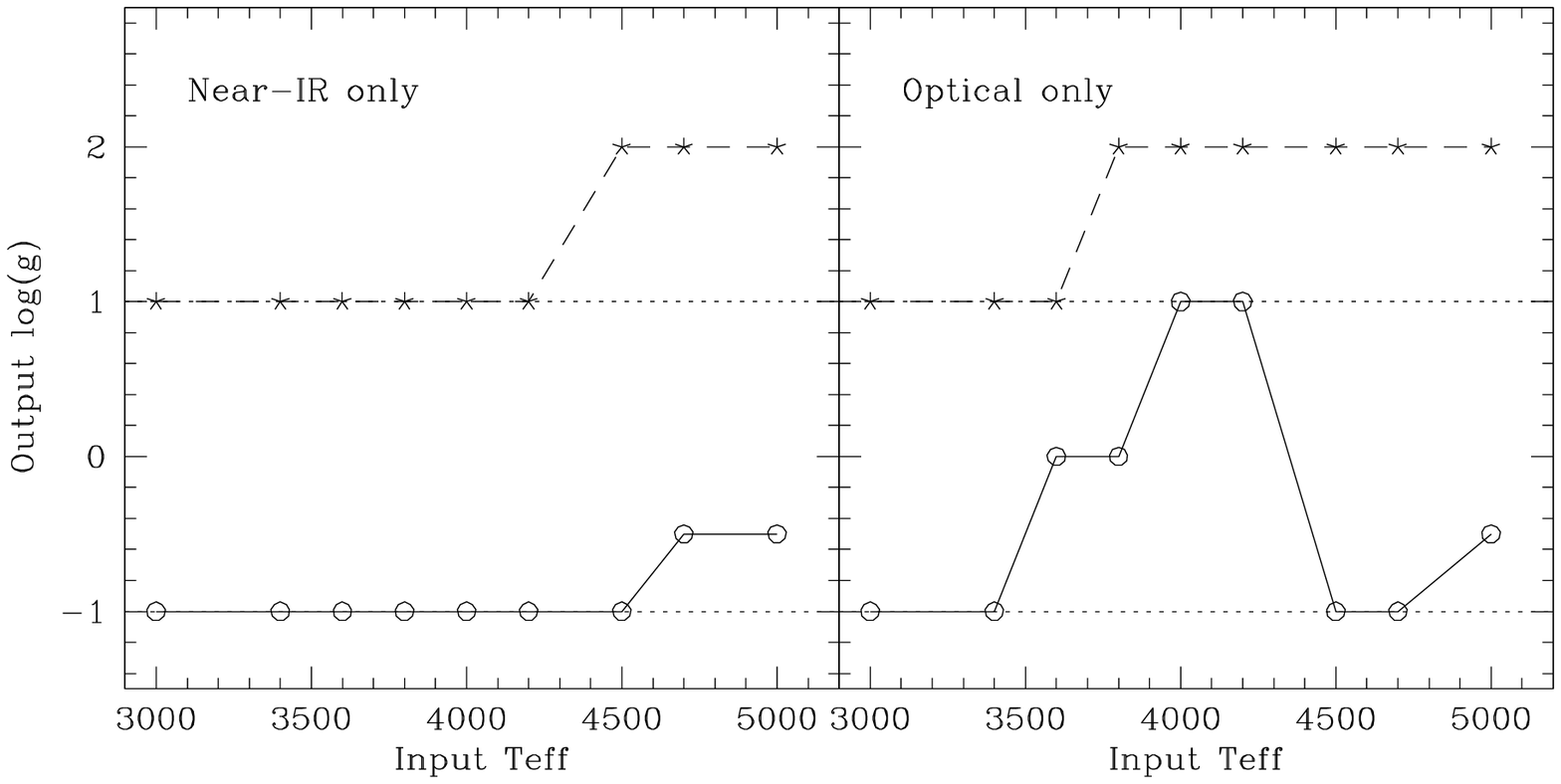}
\caption[]{Effects of the surface abundances on estimates of \Teff\ and
log($g$).  The input \Teff\ and log($g$)
refer to solar metallicity models. The output parameters to the values obtained
when fitting the solar metallicity spectra with models with RSG-specific
abundances, using the procedure described in Sect.\,\ref{method.sec}.
{\em Solid:}\ Input log($g$)=$-1$. {\em Dashed:}\ Input log($g$)=1.}
\label{modelsystematics.fig}
\end{figure}

The effects of surface abundance ratios on \Teff\ estimates
(and on derived gravities) are of larger amplitude
than those of mass, and we therefore describe them in more detail.
They can be studied by
fitting a sample of solar metallicity models  
with models with RSG-specific abundances, using the procedure described
for fits to observational data.
{\em The results depend on the wavelength range adopted in the analysis.}\
They are illustrated in Fig.\,\ref{modelsystematics.fig}. 
The amplitude of the effect is of several hundred Kelvin. If we 
call $\delta \Teff$ the difference between the input and output temperatures 
(output minus input), we find no simple linear correlation between 
$\delta \Teff$ and \Teff.  

The figure based on near-IR data (0.97--2.4\,$\mu$m, with the window 
and weight functions
of Sect.\,\ref{method.sec}) is tightly related to the behaviour of the
near-IR CN bands. Output \Teff\ values around 4400--4900\,K (depending
on gravity) are avoided because
the CN bands of those RSG-specific models are too strong. 
A similar effect is present when optical wavelengths are used 
(0.51-0.97\,$\mu$m), but it is 
combined in the low-\Teff\ range with a variety of effects due to oxides.
The difference between the optical and near-IR temperatures is
largest between 3500 and 4200\,K, where the fluxes below 0.75\,$\mu$m 
transit rapidly from being almost nil to being large. 
Eye inspection of the fits shows that the best 
fit models sometimes deviate wildly from the ``data" in the range {\em not} 
included in the fitting procedure, while over the range really used fits 
are dangerously good. When optical and near-IR spectra are
used jointly, compensations occur and the correct Teff is recovered (to
within $\pm 100$\,K) below 4200\,K. Positive offsets of up to 400\,K however
persist above this temperature for all gravities. 

The output log($g$) equals the input log($g$) below about 4300\,K 
when using near-IR data, and
is higher by one log($g$)-sampling step at higher temperatures. When
using optical data, the behaviour depends more strongly on the actual
value of the input log($g$). For high gravities, $\delta$log($g$) is positive
(one sampling step) at $\Teff > 3600\,K$. For low gravities,
$\delta$log($g$) is nil at the lowest and highest temperatures, but
peaks with a value $+2$ around 4200\,K.

We note that corresponding plots can be produced for the
``extinction"-correction A$_V$ (which accounts for colour changes in the 
analysed wavelength ranges reasonably well). The qualitative
aspects of the graphs for $\delta$A$_V$ are similar to those of $\delta \Teff$,
with a maximal amplitude of $\pm$0.6 magnitudes. 

\medskip

For comparison, we have performed a limited exploration
of the effects of metallicity (with solar scaled abundance ratios) on the 
derived temperatures. Models at log$(Z/Z_{\odot})=-0.3$ were
computed for log($g$)=1 and $-0.5$, and best fits to these were 
obtained using solar metallicity models. Plots
similar to those in Fig.\,\ref{modelsystematics.fig} were constructed. 
The effects of the change in $Z$ on the derived \Teff\ is notably {\em smaller} 
than those just described for modified abundance ratios. 
When using optical wavelengths,
the trend expected from the well known metallicity-temperature degeneracy is
found (lower temperatures are required at lower metallicity to produce similar
optical band depths). The offset varies between 100\,K (low \Teff ) and 200\,K
(high \Teff ). At near-IR 
wavelengths, the correct temperatures are recovered unchanged except for a
few deviations of $\pm 100$\,K. In both wavelength ranges, however,
gravities higher than input are derived (by one gravity bin).
For complementary discussions on metallicity effects, we refer to
Ku\v{c}inskas et al. (2006) and Levesque et al. (2006).

\subsection{\Teff\ estimates for real stars}
\label{Teffscales_data.sec}

\begin{figure}
\includegraphics[clip=,angle=270,width=0.49\textwidth]{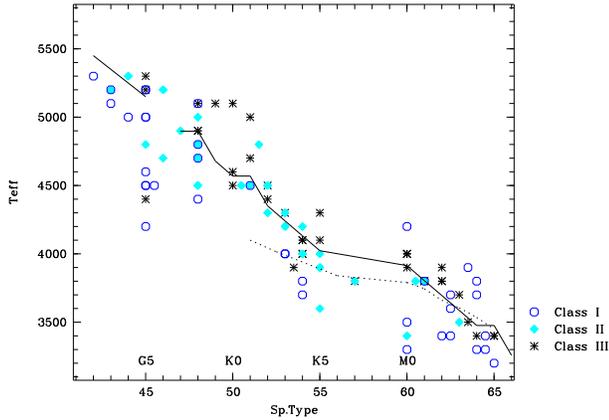}
\caption[]{Effective temperatures derived from fits to 
near-IR spectra ($\lambda>0.97\,\mu$m), 
compared with trends in the literature. RSG-specific abundances are
used for class I stars, solar abundances for classes II and III. Solid 
lines: temperature scale for giants, from van Belle et al. (1999) 
for \Teff$<$5000\,K and from Schmidt-Kaler (1982) for \Teff$>$5000\,K. 
Dotted line: temperature scale for supergiant stars from 
Levesque et al. (2005). Default spectral types from the SIMBAD 
database (operated at CDS, Strasbourg, France) are used for this figure.} 
\label{Teffscales_IR.fig}
\end{figure}


\begin{figure}
\includegraphics[clip=,angle=270,width=0.49\textwidth]{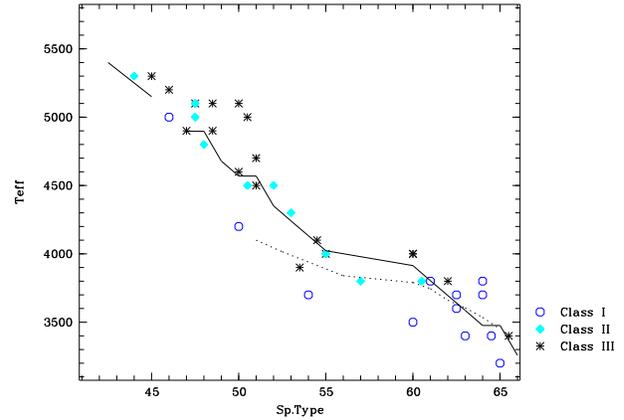}
\caption[]{Same as Fig.\,\ref{Teffscales_IR.fig}, but
using only spectral types from Keenan \& McNeil (1989)
or Buscombe (1998).
Using solar abundances moves the G6 supergiant in this figure down 200\,K, 
the K0 supergiant up 100\,K, and the M supergiants above 3600\,K up by
100 to 200\,K.}
\label{Teffscales_goodspt.fig}
\end{figure}

Figure\,\ref{Teffscales_IR.fig} compares the effective temperatures derived
in this paper from near-IR spectra with temperature scales from the literature. 
For giants, the plotted reference scale
(below 5000\,K) is based on angular diameter measurements (van
Belle et al. 1999).
The number of red supergiants with angular diameter measurements is 
small. For supergiants, we therefore show the scale recently obtained
from fits of \Marcs\ model atmosphere spectra to optical spectra by 
Levesque et al. (2005). The agreement is good, but the scatter is large.

As a sanity check, we may restrict our data sample to 
stars with {\em optical} spectra and discard near-IR wavelengths 
($\lambda > 1\,\mu$m) in the fitting procedure. 
This provides temperatures that are {\em a priori} more
directly related to spectral types.
In addition, we keep only stars with 
MK spectral types from Keenan \& McNeil (1989) or Buscombe (1998),
and with small variability (according to the Simbad database
information). {\em Using solar abundance models} 
for direct comparison with the 
results of Levesque et al. (2005), we find that 8 of the 9 stars in
the subsample have estimated temperatures within less than 50\,K of 
the reference relations. Most of the stars in the subsample are supergiants
and all are of type K5 or later. Thus, in this range of parameters,
there is no indication of a systematic 
difference between the temperatures derived from optical spectra
using the new \Phoenix\ models or the \Marcs\ models of 
Levesque et al. (2005).
 
To illustrate what fraction of the scatter in Fig.\,\ref{Teffscales_IR.fig}
may be due to spectral classification errors, 
Fig.\,\ref{Teffscales_goodspt.fig} reproduces the graph using only
MK spectral types from Keenan \& McNeil (1989) or Buscombe (1998),
when available. 
A considerable scatter remains. Some of it is due to the real scatter
in the properties of the stars (surface abundances, gravity, unknown 
variability).
For supergiants in particular, and especially at low temperatures,
the scatter also reflects the large 
intrinsic uncertainties associated with the relatively poor quality of
the model fits. We expect the spread to shrink once 
models with a wider range of parameters (surface abundances, micro-turbulent
velocities) will have been computed.

We have also examined the diagrams of estimated \Teff\ vs. spectral type
obtained when any available optical data is included in the fitting 
procedure. They are similar to those described above. Individual
stars are moved by up to 200\,K, but no systematic trend can be clearly
identified (because the stars that move most are also those for which the
fits are poorest). Despite the added difficulty of fitting a broader
wavelength range, the final dispersion is not significantly enhanced.

\section{Conclusions}

We have presented two grids of \Phoenix\ models for the spectra of
luminous cool stars, one at solar metallicity, the other with RSG-specific
surface abundances. We have described the properties of these models
and compared them with observations, with a focus on the molecular
features found in near-IR spectra at resolution 
$\lambda/\Delta \lambda \simeq 1000$. At these 
wavelengths, red giants and supergiants dominate the integrated light of 
stellar populations.
Our main conclusions are the following.
\begin{itemize}
\item Models must be computed with a wavelength sampling step of about 
0.1\,\AA\ in order to reproduce the low resolution near-IR spectra 
adequately. 
\item The solar metallicity models provide a very good representation of 
empirical spectra of giants of class III and of a large fraction of the
luminous giants of class II. As expected, RSG-specific abundances are found
inadequate for the modelling of the bulk of the giant stars (they are
rejected because they provide poorer fits and lead to a
zone of avoidance in the derived \Teff -distribution).
RSG-specific abundances are favoured for
some class II giants, which may have suffered mixing in excess of standard
first dredge-up. 
\item Red supergiant spectra of spectral types G and K, and of
luminosity class Ib (sometimes also Iab) can be reproduced reasonably well.
Serious disagreements remain in the case of very luminous (Ia and some Iab) 
and of very cool supergiants (type M).  
RSG-specific abundances tend to improve the fits to strong
CN bands, although the global effect on the fit quality is not as
spectacular as one might have hoped. However, 
changing the surface abundance ratios has a significant 
impact on the derived effective temperatures (the 
effect is larger than that found when moving from $0.5\,Z_{\odot}$ to 
$Z_{\odot}$). Therefore,
it will remain necessary to account for this effect of stellar evolution
in future model grids.  
\item While it is easy (relatively) 
to produce good fits to the spectra of either the J,
{\em or} the H, {\em or} the K band spectra of luminous cool stars, 
it remains more difficult to reproduce all their optical and near-IR 
molecular bands simultaneously. As a result, estimated stellar parameters 
(\Teff, log($g$), A$_V$) depend on the spectral range of the analysis.
The effects of changes in the surface abundances on these parameters
also depend on the wavelengths under study.
\item The \Teff\ scales derived from the comparison of a collection
of near-IR stellar spectra ($1-2.4\,\mu$m)
with models are generally consistent with previous scales, 
albeit with considerable scatter. For cool red supergiants, the 
current uncertainties on individual estimated \Teff\ values frequently 
exceed $\pm 100$\,K.
\end{itemize}
About 20\,\% of the analysed red supergiant spectra have such strong CN
bands that they call for models with high micro-turbulent velocities, 
and/or even more surface nitrogen than we 
have considered, and/or for gravities lower than log($g$)=$-1$. 
The coolest of these are variable, and variability may contribute
to the building of an extended atmosphere with low effective gravities. 
Large micro-turbulent velocities have been derived for a 
number of red supergiants in the past, and our first calculations confirm
that increasing this parameter will help reproducing the spectra
of type Ia supergiants. In particular, a better agreement with observations 
is expected for the ratio between the first and second overtone
CO band strengths.  A grid of models is currently being
calculated. Somewhat higher nitrogen abundances than we have explored
are expected to exist in nature, for instance when stellar 
rotation increases internal mixing.  Because
low resolution near-IR spectra of red supergiants are 
relatively easy to acquire, their comparison with models at the 
specific abundances predicted by stellar tracks with rotation will
provide interesting tests of stellar evolution theory. 
Considering stars
with lower gravities is a more challenging modelling task, 
as they will develop strong winds. In addition, the winds may
be loaded with dust. Since winds are 
a well known empirical property of many red supergiants, the development
of models that include winds is a necessity of the future. 
  
%

%
%
%
%
%
%

\begin{acknowledgements}

PHH was supported in part by the P\^ole Scientifique de Mod\'elisation
Num\'erique at ENS-Lyon and by Universit\'e Louis Pasteur at Strasbourg 
Observatory.  Some of the calculations presented here were  
performed
at the H\"ochstleistungs Rechenzentrum Nord (HLRN), and at the  
National Energy
Research Supercomputer Center (NERSC), supported by the U.S. DOE, and  
at the
computer clusters of the Hamburger Sternwarte, supported by the DFG  
and the
State of Hamburg.  We thank all these institutions for a generous
allocation of computer time.
We thank C. Charbonnel for insightful discussions of 
aspects of this work.
This research has made use of the SIMBAD database and the VIZIER service, 
operated at CDS, Strasbourg, France. It uses data (in preparation 
for publication) acquired using the NASA Infrared Telescope 
Facility, Hawaii, USA, and the 2.3m Telescope of the Australian National
University, Siding Spring, Australia. 

\end{acknowledgements}


\appendix


\begin{thebibliography}{}
\bibitem{AndersGrev89} Anders, E., Grevesse, N. 1989, 
 Geochim. Cosmochim. Acta 53, 197
\bibitem{Aspletal05} Asplund, M., Grevesse, N., Sauval, A.J. 2006, in
  Cosmic Abundances as Records of Stellar Evolution and Nucleosynthesis,
  T.G. Barnes, III, \& F.N. Bash (eds.), ASP Conf. Ser. 336, 25
\bibitem{Baldwinetal73} Baldwin, J.R., Frogel, J.A., Persson, S.E. 1973,
  ApJ 184, 427
\bibitem{BeckIben79} Becker, S.A., \& Iben, I., Jr. 1979, ApJ 232, 831
\bibitem{Beeretal72} Beer, R., Hutchison, R.B., Norton, R.H., Lambert, D.L. 
  1972, ApJ 172, 89
\bibitem{BessBrett88} Bessell, M.S., Brett, J.M. 1988, PASP 100, 1134
\bibitem{Bessell_etal89} Bessell, M.S., Brett, J.M., Scholz, M., \& Wood, P.R.
 1989, A\&AS 77, 1
\bibitem{Bessell_etal91} Bessell, M.S., Brett, J.M., Scholz, M., \& Wood, P.R.
 1991, A\&AS 89, 335
  %
\bibitem{BoothSack99} Boothroyd, A.I., \& Sackmann, I.-J. 1999, ApJ 510, 232
\bibitem{Bressanetal93} Bressan, A., Fagotto, F., Bertelli, G., Choisi, C. 1993,
  A\&AS 100, 647
\bibitem{Buscombe98} Buscombe, 1998, Illinois catalogue of MK spectral 
  classifications (version 13), Northwestern Univ. Evanston, Illinois.
\bibitem{Cardelli_etal89} Cardelli, J.A., Clayton, G.C. \& Mathis, J.S. 1989,
  ApJ 345, 245
\bibitem{Charbonnel94} Charbonnel, C. 1994, A\&A 282, 811
\bibitem{CharbdoNacim98} Charbonnel, C. \& do Nascimento, J.D., Jr. 1998, A\&A
  336, 915
\bibitem{Charbetal96} Charbonnel C., Meynet, G., Maeder, A., Schaerer, D. 1996,
 A\&AS 115, 339
\bibitem{Decinetal04} Decin, L., Shkedy, Z., Molenberghs, G., Aerts, M., 
  Aerts, C. 2004, A\&A 421, 281
\bibitem{Frogeletal78} Frogel, J.A., Persson, S.E., Aaronson, M.,
  Matthews, K. 1978, ApJ 220, 75
\bibitem{Girardietal00} Girardi, L., Bressan, A., Bertelli, G., Chiosi, C. 2000,
  A\&AS 141, 371
\bibitem{GoorCOa94} Goorvitch, D., \& Chackerian, C. 1994, ApJS 91, 483
\bibitem{GoorCOb94} Goorvitch, D., \& Chackerian, C. 1994, ApJS 92, 311
\bibitem{Grevesse91} Grevesse, N. 1991, in Evolution of Stars: the Photospheric
  Abundance Connection, G. Michaud \& A.V. Tutukov (Eds.), Kluwer (Dordrecht), 
  IAU Symp. 145, 63
\bibitem{GrevNoels93} Grevesse, N., \& Noels, A. 1993, in Origin and Evolution
  of the Elements, N. Prantos, E. Vangioni-Flam \& M. Cass\'e (Eds.),
  Cambridge Univ. Press, 14
\bibitem{Iben64} Iben, I., Jr. 1964, ApJ 140, 1631
\bibitem{Iben66} Iben, I., 1966, ApJ 143, 516
\bibitem{KMcN89} Keenan, P.C., McNeil, R.C. 1989, ApJSS 71, 245
\bibitem{Kucinetal05} Ku\v{c}inskas, A., Hauschildt, P.H., Ludwig, H.-G., 
  Brott, I., Vansevi\v{c}ius, V., Lindegren, L., Tanab\'e, T., Allard, F. 2005,
  A\&A, 442, 281 
\bibitem{Kucinetal06} Ku\v{c}inskas, A., Hauschildt, P.H., Brott, I.,
  Vansevi\v{c}ius, V., Lindegren, L., Tanab\'e, T., Allard, F. 2006,
  A\&A 452, 1021
\bibitem{LanconWood00} Lan\c{c}on, A., Wood, P.R. 2000, A\&AS 146, 217
\bibitem{Lanconetal03} Lan\c{c}on, A., Smith, L.J., Gallagher, J.S. et al.
  2003, in Extragalactic Globular Clusters and their Host Galaxies
  (IAU JD 6, ed. T.\,Bridges), Highlights of Astronomy 13, 
  ed. O.\,Engvold, in press
\bibitem{Lanconetal06} Lan\c{c}on, A., Wood, P.R., Ladjal, D., Mouhcine, M., 
  Gallagher, J.S., Smith, L.J., Vacca, B., F\"orster Schreiber, N.,
  de Grijs, R., O'Connell, R. 2007, in preparation
\bibitem{Levesqetal05} Levesque, E.M., Massey, P., Olsen, K.A.G., Plez, B.,
  Josselin, E., Maeder, A., Meynet, G. 2005, ApJ 628, 973
\bibitem{Levesquetal06} Levesque, E.M., Massey, P., Olsen, K.A.G., Plez, B.,
  Meynet, G., Maeder, A. 2006, ApJ 645, 1102
\bibitem{LoidlLJ01} Loidl, R., Lan\c{c}on, A., J\/orgensen, U.G. 2001,
 A\&A 371, 1065
\bibitem{Maeder81} Maeder, A. 1981, A\&A 101, 385
\bibitem{MaedMeyn01VII} Maeder, A., \& Meynet, G. 2001, A\&A 373, 555
\bibitem{Martinsetal05} Martins, L.P., Gonz\'alez Delgado, R.M., Leitherer, C.,
  Cervi\~no, M., Hauschildt, P.H. 2005, MNRAS 358, 49
\bibitem{McGregor87} McGregor, P.J. 1987, ApJ 312, 195
\bibitem{McWillLambert84} McWilliam, A., \& Lambert, D.L. 1984, PASP 96, 882
\bibitem{MeynetMaed00V} Meynet, G., \& Maeder, A. 2000, A\&A 361, 101
\bibitem{MeynetMaed03X} Meynet, G., \& Maeder, A. 2003, A\&A 404, 975
\bibitem{Meynetetal94} Meynet, G., Maeder, A., Schaller, G., Schaerer, D.,
  Charbonnel, C. 1994, A\&AS 103, 97
\bibitem{Origliaetal93} Origlia, L., Moorwood, A.F.M., Oliva, E. 1993, 
   A\&A 280, 536
\bibitem{Origliaetal97} Origlia, L., Ferraro, F.R., Fusi Pecci, F., Oliva, E.
  1997, A\&A 321, 859
\bibitem{Schalleretal92} Schaller, G., Schaerer, D., Meynet, G.,
  Maeder, A. 1992, A\&AS 96, 269
\bibitem{Schmidt-Kaler82} Schmidt-Kaler 1982, in Astrophysical Data\,: Planets
 \& Stars (1992), Lang N. (Ed.), Springer (New York)
  %
\bibitem{Schwenke98} Schwenke, D.W. 1998, 
  in Chemistry \& Physics of Molecules and Grains in Space. 
  Faraday Discussions No. 109 (Royal Soc. of Chemistry, London) p.321.
\bibitem{SmithLamb85} Smith, V.V., \& Lambert, D.L. 1985, ApJ 294, 326
\bibitem{TrundleLen05} Trundle, C., \& Lennon, D.J. 2005, A\&A 434, 677
\bibitem{Tsuji76} Tsuji, T. 1976, PASJ 28, 543
\bibitem{vBetal99} van Belle, G.T., Lane, B.F., Thompson, R.R. et al. 1999,
  AJ 117, 521
\bibitem{Vanholetal06} Vanhollebeke, E., Blommaert, J.A.D.L., 
  Schultheis, M., Aringer, B., Lan\c{c}on, A. 2006, A\&A 455, 645
\bibitem{WhiteWing78} White, N.M., Wing, R.F. 1978, ApJ 222, 209
\bibitem{WingSpin70} Wing, R.F., \& Spinrad, H. 1970, ApJ 159, 973
\end{thebibliography}
\end{document}